\documentclass[useAMS,usenatbib]{mn2e}
\usepackage{graphicx}
\usepackage{epsfig}
\usepackage{subfigure}
\usepackage{color}
\usepackage{xcolor}
\usepackage{multirow}
\usepackage{threeparttable}
\usepackage{natbib}
\usepackage{amsmath,ulem,upgreek,amssymb}
\usepackage[colorlinks=true,linkcolor=blue,citecolor=blue]{hyperref} % Hyperlink capabilities
\title[A383: Constraining the galaxy mass content in cluster cores]{Constraining the galaxy mass content in the core of A383 using velocity dispersion measurements for individual cluster members}

\author[A. Monna et al.]
{\parbox{\textwidth}{ A. Monna$^{1,2}$\thanks{E-mail:
amonna@usm.uni-muenchen.de},
S. Seitz$^{1,2}$,
A. Zitrin$^{3,4}$,
M. J. Geller$^{5}$,
C. Grillo$^{6}$,
A. Mercurio$^{7}$,
N. Greisel$^{1,2}$,
A. Halkola$^{}$,
S. H. Suyu$^{8}$,
M. Postman$^{9}$,
P. Rosati$^{10}$,
I. Balestra$^{2,11}$,
A. Biviano$^{11}$,
D. Coe$^{9}$,
D. G. Fabricant$^{5}$,
H. S. Hwang$^{12}$,
A. Koekemoer$^{9}$}\vspace{0.4cm}\\
\parbox{\textwidth}{$^{1}$University Observatory Munich, Scheinerstrasse 1, 81679 Munich, Germany\\
$^{2}$Max Planck Institute for Extraterrestrial Physics, Giessenbachstrasse, 85748 Garching, Germany\\
$^{3}$Cahill Center for Astronomy and Astrophysics, California Institute of Technology, MS 249-17, Pasadena, CA 91125, USA\\
$^{4}$Hubble Fellow\\
$^{5}$Harvad-Smithsonian Astrophysical Observatory, 60 Garden St., Cambridge, MA 02138\\
$^{6}$Dark Cosmology Centre, Niels Bohr Institute, University of Copenhagen, Juliane Maries Vej 30, 2100 Copenhagen, Denmark\\
$^{7}$INAF/Osservatorio Astronomico di Capodimonte, Via Moiariello 16, I-80131 Napoli, Italy\\
$^{8}$Institute of Astronomy and Astrophysics, Academia Sinica, P.O. Box 23-141, Taipei 10617, Taiwan\\
$^{9}$Space Telescope Science Institute, 3700 San Martin Drive, Baltimore, MD 21208, USA\\
$^{10}$Dipartimento di Fisica e Scienze della Terra, Univ. degli Studi di Ferrara, via Saragat 1, I-44122, Ferrara, Italy\\
$^{11}$INAF-Osservatorio Astronomico di Trieste, via G.B. Tiepolo 11, 34143 Trieste, Italy\\
$^{12}$Korea Institute for Advanced Study, 85 Hoegiro, Dongdaemun-gu, Seoul 130-722, Republic of Korea\\
}}
\begin{document}
 \date{}

% \pagerange{\pageref{firstpage}--\pageref{lastpage}} \pubyear{2010}

\maketitle

\label{firstpage}

\begin{abstract}
%229 words now
We use  velocity dispersion measurements of 21 individual cluster members in the core of Abell 383, obtained with MMT Hectospec, to 
separate the galaxy and the smooth dark halo (DH) lensing contributions.
While lensing usually constrains the overall, projected mass density, the innovative use of  velocity dispersion measurements as a proxy for masses of individual cluster members  breaks inherent degeneracies and allows us to (a) refine the constraints on single galaxy masses and on the galaxy mass-to-light scaling relation and, as a result, (b)  refine the constraints on the DM-only map, a high-end goal of lens modelling. The knowledge of cluster member  velocity dispersions improves the fit by 17\% in terms of the image reproduction $\chi^2$, or 20\% in terms of the $rms$. 
The constraints on the mass parameters improve by $\sim10\%$ for the DH,
while for the galaxy component, they are refined  correspondingly by $\sim50\%$, including the galaxy halo truncation radius.
For an $L^{*}$ galaxy with $M^{*}_{B}=-20.96$, for example, we obtain best fitting truncation radius $r_{\rm tr}^{*}=20.5^{+9.6}_{-6.7}$\, kpc and  velocity dispersion $\sigma_{*}=324\pm17$\, km/s.
Moreover,  by performing the surface brightness reconstruction of the southern giant arc, we improve the constraints on  $r_{\rm tr}$ of two nearby cluster members, which have measured velocity dispersions, by more than $\sim30\%$.
We estimate the stripped mass for these two galaxies, getting results that are consistent with numerical simulations. 
In the future, we plan to apply this analysis to other galaxy clusters for which velocity dispersions of member galaxies are available.
\end{abstract}

\begin{keywords}
dark matter, galaxy cluster, galaxy halos, gravitational lensing.
\end{keywords}

\section{Introduction}
Gravitational lensing and its modelling represent reliable and important tools to map the mass distribution of structures in the Universe, from galaxies through galaxy groups and clusters, up to the large-scale structure \citep[e.g.,][]{Schneider2003,Bartelmann2010,Kneib2011}. 
One of the main motivations of using lensing is its ability to map the total projected mass density of the lens and thus shed light on the distribution and properties of the otherwise invisible dark matter (DM).\\ 
Modelling of gravitational lensing is usually performed in two  ways. 
The first, often dubbed 'non-parametric', elegantly makes no prior assumptions on the underlying mass distribution, but due to the typical low number of constraints usually yields a low-resolution result that lacks predictive power \citep[see,][]{Abdelsalam1998,Diego2005,Coe2008}. 
Alternatively, 'parametric' mass models exploits prior knowledge or assumptions regarding the general form of the underlying mass distribution. The mass  parameters are obtained by producing many mass models, each with a different set of parameter values, and looking for the solution which best reproduces the observations. 
Despite their model dependence, these methods allow for a very high spatial resolution, and typically exhibit high predictive power to reproduce additional constraints such as multiple images not used as inputs \citep[see e.g.,][]{Jullo2007,Zitrin2009b, Monna2014, Grillo2014}.
In the case of galaxy clusters acting as lenses, the cluster DM component usually follows descriptions obtained from numerical simulations, such as an elliptical NFW halo or alike.
The galaxy mass component of the cluster is given by the combination of all the cluster member masses which are typically modelled as power-law profiles, isothermal spheres or their variants \citep[e.g., see][]{Nat1997,Kneib2011}. 
The combination of the baryonic and DM components yields the total projected surface mass density, which is the quantity probed in the lensing analysis. 
In that respect, it is difficult to properly separate the baryonic and DM galaxy components, as lensing probes only their joint contribution and  degeneracies exist between the different parameters which could explain the same set of constraints. 
To infer the masses of the galaxies directly from the light, typically, luminosity-velocity dispersion-mass scaling relations are used.
Physical properties of elliptical galaxies are globally well described by power-law relations which relate them to their observed luminosity, both for galaxies in field and in clusters. 
The Fundamental Plane \citep[see][]{Bender1992,Dressler1987,Djorgovski1987, Faber1987} gives the relation 
between  effective radius $r_e$, central velocity dispersion $\sigma_0$ and mean surface brightness $I_e$ within $r_e$ of elliptical galaxies. 
The central velocity dispersion $\sigma_0$ is related to the galaxy luminosity $L_e$ through the Faber-Jackson relation ($L_e\propto\sigma_0^\alpha$) (Faber \& Jackson 1976). However, it has been shown that bright galaxies, like the brightest cluster galaxies (BCGs), can  deviate substantially from the scaling relation \citep[see][]{Vonderlinden2007,Postman2012b,Kormendy2013}. \\
In \cite{Eichner2013} we investigated the halo properties of the cluster members of MACS1206.2-0847 through strong lensing analysis. 
We broke the degeneracy between the halo velocity dispersion $\sigma$ and size $r_{\rm tr}$, improving the constraints on the $\sigma-r_{\rm tr}$ relation through the surface brightness reconstruction of the giant arc in the core of the cluster.
However, the large scatter in the Fundamental Plane (or, Faber-Jackson relation) inherently introduces modelling biases in lensing analyses which inevitably assume an analytic scaling relation for the M-L-$\sigma$ planes. 
Direct velocity dispersion measurements of  galaxies (typically, elliptical cluster members) allow a direct estimate of their enclosed mass, through the virial theorem that reduces to $\rho(r)=\frac{\sigma^2}{2\pi Gr}$ for an isothermal sphere, for example. 
These mass estimates can be used individually for each lens galaxy instead of applying an idealised analytic scaling relation. 
This will especially be significant for bright and massive cluster galaxies governing the lens, i.e. galaxies within, or close to, the critical curves, as these affect the lensing properties the most.
For that reason, we have embarked on an innovative project using Hectospec on the Multiple Mirror Telescope (MMT) to measure the velocity dispersion of individual cluster members in various cluster lenses. 
We aim to obtain an independent measure of the mass of each relevant cluster galaxy, so that internal degeneracies can be broken and the constraints on the DM-only component improved. 
We present here the first case-study we perform using these velocity dispersion data, analysing the strong-lensing features in the galaxy cluster Abell 383 (hereafter A383) at $z_{\mathrm{cl}}=0.189$, while examining the extent of improvement obtained by using these additional mass proxies. 
\\
The mass distribution of A383  has been previously traced through gravitational lensing analyses \citep[see][]{Smith2001,Smith2005,Sand2004}, also combined with dynamical analyses \citep[see][]{Sand2008}.
\citet{Newman2011} combined strong and weak lensing  analyses with galaxy kinematics and X-Ray data to trace the mass distribution of the cluster out to 1.5 Mpc. 
They disentangled the baryonic and dark matter components in the inner region of the cluster, finding a shallow slope $\beta$ for the density profile $\rho\propto r^{-\beta}$ of the dark matter on small scales.
\citet{Geller2014} presented a detailed dynamical analysis of A383 using 2360 new redshift measurements of galaxies in the region around the cluster. 
They traced the cluster mass distribution up to about 7~Mpc from the cluster centre, obtaining results that are in good agreement with mass profiles derived from weak lensing analyses, in particular at radial distances larger than $R_{200}$. 
\citet{Zitrin2011} performed a detailed strong lensing reconstruction of the cluster using the well known giant arcs and several newly identified lensed systems using the deep 16-band HST photometric dataset from the CLASH survey \citep{Postman2012a}. 
They used 9 lensed systems with a total of 27 multiple images to measure in detail the total mass distribution and profile in the cluster core. 
 
In the work presented here we perform an accurate strong lensing analysis of A383 using velocity dispersion measurements  for several cluster members as additional constraints. 
We investigate the impact of such information on the accuracy of the lensing reconstruction, on the constraints for the individual galaxy masses and on the global $\sigma-L$ relation. 
In addition, we perform the surface brightness reconstruction of the southern tangential  giant arc lensed between several cluster members to set stronger constraints on the mass profiles of these individual galaxies and directly measure their size.\\

The paper is organised as follows.
In Section~\ref{sec:dataset} we describe the photometric and spectroscopic dataset. 
In Section~\ref{sec:catalogs} we present the photometric catalogues and the cluster member selection. 
In Section~\ref{sec:lensing} we describe the strong lensing analysis, the mass components included in the mass model and the lensed systems used as constraints. 
In Section~\ref{sec:pointlike_model} we present the results of the  strong lensing analyses performed using as constraints the observed positions of lensed images. 
In Section~\ref{sec:extended_image} we perform the surface brightness reconstruction of the southern giant arc to refine the constraints on the mass profile of cluster members close to the arc.  
Summary and conclusions are given in Section~\ref{sec:conclusions}.
Throughout the paper we assume a cosmological model with Hubble constant $H_0 = 70$ km s$^{-1}$ Mpc$^{-1}$ and density parameters  $\Omega_{\rm m} = 0.3$ and $\Omega_{\Lambda}=0.7$. 
Magnitudes are given in the AB system.

\section{Photometric and Spectroscopic Dataset} 
\label{sec:dataset}
 \begin{table}
\caption{Photometric dataset summary: column (1) filters, column (2) HST instrument,  column (3) total exposure time in seconds, column (4) $5\sigma$ magnitude depth within $0.6\arcsec$ aperture (see text).}
\centering
\footnotesize
\begin{tabular}{|c|c|c|c|}
\hline
\hline
Filter&Instrument& Exposure time [s]& 5$\sigma$ Depth\\
%      &             &   \scriptsize{A383}\quad  &\quad \scriptsize{RXJ2129}&\scriptsize{A383 } & \scriptsize{RXJ2129}\\
      \hline
F225W &   WFC3/UVIS &   7343 &    25.76\\
F275W &   WFC3/UVIS &   7344 &    25.84\\
F336W &   WFC3/UVIS &   4868 &    26.06\\
F390W &   WFC3/UVIS &   4868 &    26.68\\
F435W &  ACS/WFC   &   4250 &     26.47  \\    
F475W &  ACS/WFC   &   4128 &     26.81  \\  
F606W &  ACS/WFC   &   4210 &     27.06    \\
F625W &  ACS/WFC   &   4128 &     26.55    \\
F775W &  ACS/WFC   &   4084 &     26.46    \\
F814W &  ACS/WFC   &   8486 &      26.79  \\
F850LP&  ACS/WFC   &  8428 &       25.93   \\
F105W &  WFC3/IR   & 3620 &     26.81        \\
F110W &  WFC3/IR   & 2515 &     27.09        \\
F125W &  WFC3/IR   & 3320 &     26.68        \\
F140W &  WFC3/IR   & 2412 &     26.80        \\
F160W &  WFC3/IR   & 5935 &    26.81         \\
\hline                  
\end{tabular}
\label{tab:phot}
\end{table}
As part of the CLASH survey, A383 was observed  (between November 2010 and March 2011) in 16 filters covering the UV, optical and NIR range with the HST Advanced Camera for Surveys (ACS) and  the  HST Wide Field Camera 3 (WFC3) with its UVIS and IR cameras. 
The photometric dataset\footnote{available at http://archive.stsci.edu/prepds/clash/} is composed of HST mosaic drizzled 65mas/pixel images generated with the \texttt{Mosaicdrizzle} pipeline \citep[see][]{Koekemoer2011}. 
They cover a field of view (FOV) of $\sim 2.7'\times2.7'$ in the UVIS filters, $\sim 3.4'\times3.4'$ in the ACS and $\sim2'\times2'$ in the WFC3IR images, centred on the cluster core.  
In Tab.~\ref{tab:phot}  we list the filters, observing times and depths of the photometric data.
For each band, we estimate the detection limit by measuring the fluxes within $3000$ random apertures of $0.6\arcsec$ diameter within the image FOV. 
We generate multi-band photometric catalogues of  fluxes extracted within $0.6\arcsec$  diameter aperture
using \texttt{SExtractor} 2.5.0 \citep{Bertin1996} in dual image mode. 
As detection image we use the weighted sum of all WFC3IR images. \\
The cluster is also part of the CLASH-VLT Large Program 186.A-0798 (P.I. Rosati P.). 
This survey aims to follow up the 14 southern clusters of the HST CLASH survey and provide hundreds of redshifts for cluster members, lensing features, high-z magnified galaxies and supernovae hosts. 
We use preliminary spectroscopic results from the first VIMOS observations of the cluster, taken between October 2010 and March 2011. 
The observations were performed with the LR-Blue and MR-Red grisms of the VIMOS spectrograph, providing a FOV of $\sim25\arcmin$.   
These spectroscopic data result in $\sim1000$ redshift measurements in the field of the cluster. 
They confirm 13 cluster members in the core ($r<1.5'$) of A383 and provide spectroscopic redshift measurements for $4$ multiply lensed systems. 
One of these strongly lensed systems is a double imaged $z\sim6$ source identified in the HST CLASH data and  presented in \citet{Richard2011}. 
The cluster VLT/VIMOS observations have been completed in 2014, and this complete spectroscopic dataset will be published in Rosati et al. (in prep.). \\ 

 \begin{figure*}
 \centering
 \includegraphics[width=16cm]{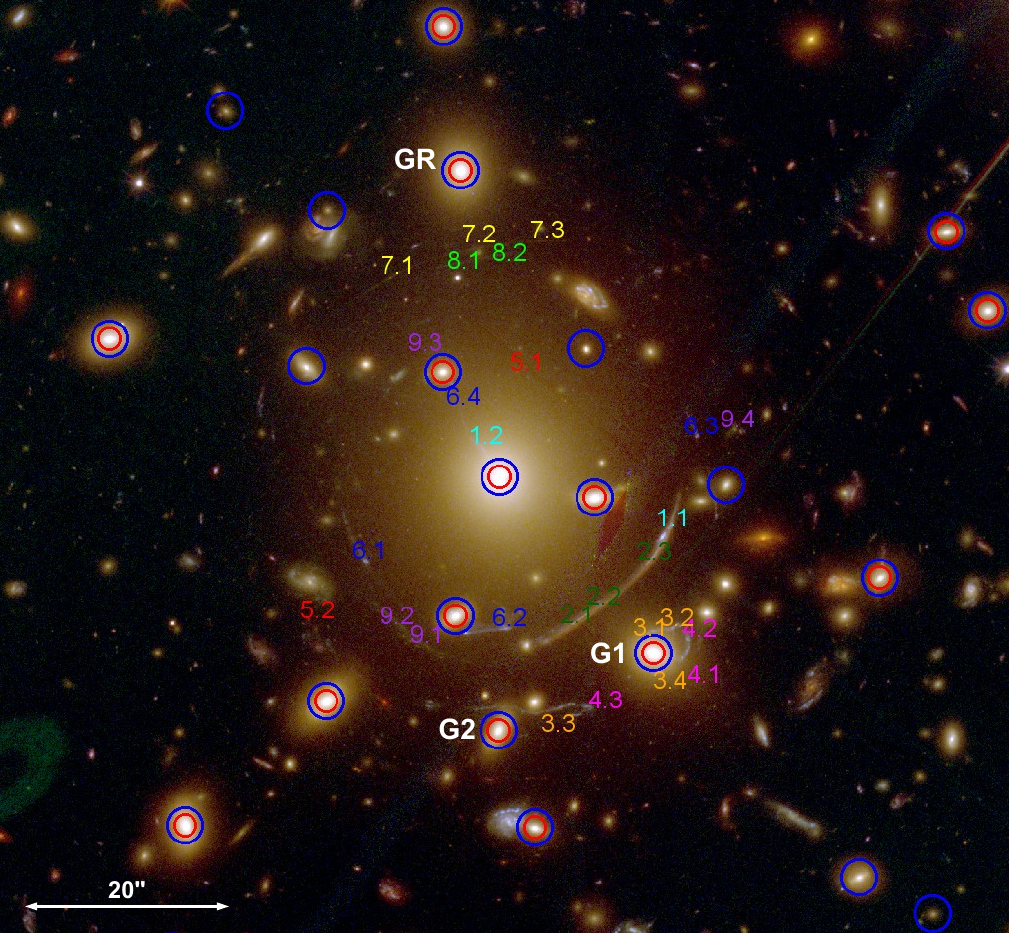}
 \caption{\small $\rm1.5'\times1.5'$ HST colour composite image of A383 core: Blue=F435W+F475W; Green=F606W+F625W+F775W+F814W+F850LP; Red=F105W+F110W+F140W+F160W. 
 Blue circles label the spectroscopically confirmed cluster members in the FOV shown, while red circles label the galaxies with measured velocity dispersion. We label with ``GR'' the galaxy used as reference for the luminosity scaling relations, and with ``G1'' and ``G2'' the two galaxies we are modelling individually (see Sec.~\ref{sec:extended_image}). 
 The 9 multiply lensed systems used in the strong lensing analysis (see Sec.~\ref{sec:lensing}) are labelled as in \citet{Zitrin2011}. }
         \label{fig:a383rgb}
 \end{figure*}

In addition, we use the sample of  galaxies observed within the Hectospec redshift survey \cite{Geller2014}.
\citet{Geller2014} used  Hectospec \citep{Fabricant2005} mounted on the 6.5-meter
MMT to measure 2360 redshifts within 50$^\prime$ of the centre of A383. 
Hectospec is a multi-object fiber-fed spectrograph with 300 fibers with an aperture of $1.5$\arcsec, deployable over a circular field-of-view with a diameter of 1$^\circ$. 
The spectra cover the wavelength range 3500 - 9150\AA.

During the pipeline processing based on the IRAF cross-correlation package \texttt{rvsao} \citep{Kurtz1998}, spectral fits receive a quality flag ``Q'' for high-quality redshifts, ``?" for marginal cases, and ``X'' for poor fits. 
All 2360 redshifts published by Geller et al (2014) have quality Q.

To derive a velocity dispersion for a galaxy we follow the procedure outlined by \citet{Fabricant2013}. 
We use an IDL-based software package, \texttt{ULySS}, developed by \citet{Koleva2009} to perform direct fitting of Hectospec spectra over the interval 4100 to 5500 Angstroms. 
The effective resolution of the Hectospec spectra in this interval is 5.0-5.5 \AA.  
There are 70 galaxies in the entire \citet{Geller2014} A383 survey that have velocity dispersions with errors $ < 25$ km s$^{-1}$ and a spectral fit with reduced $\chi^2 < 1.25$. 
These spectra have a median signal-to-noise of 9.5 over the wavelength 4000-4500 \AA. 
Among these objects, 21 are in the core of the cluster and we report them here in Table~\ref{tab:vel_gal}.

\cite{Jorgensen1995}  empirically show that the stellar velocity dispersion $\sigma_{obs}$ observed with fibers  and the central stellar velocity dispersion $\sigma_{sp}$ are related by:
\begin{equation}
 \sigma_{\rm sp}=\sigma_{\rm obs}\left(\frac{\rm R_{eff}}{8\times d/2}\right)^{-0.04}
\label{eq:sigma}
 \end{equation}
where $\rm R_{eff}$ is the galaxy effective radius and $d$ is the fiber aperture. 
We estimate the effective radii of cluster members using \texttt{GALFIT} \citep{Peng2010}, fitting de Vaucouleurs profiles to  the 2D surface brightness distribution of the galaxies in the HST/F814W filter. 
We then correct the central velocity dispersion for our cluster members according to Eq.~\ref{eq:sigma}.\\ 
In Table\,\ref{tab:vel_gal} we provide the coordinates, spectroscopic redshift $z_{\mathrm{sp}}$ and $\sigma_{\rm sp}$ for the sample of cluster members confirmed in the core of the cluster.   
 \begin{table}
 \caption{List of cluster members with measured spectroscopic redshift from the Hectospec and VIMOS/VLT surveys. Col.\,1 ID; Col.\,2-3 Ra and Dec; Col.\,4 spectroscopic redshift; Col.\,5 measured velocity dispersion corrected according to Eq.\,\ref{eq:sigma}; Col.\,6 Effective radius.}
 \begin{threeparttable}
\label{tab:vel_gal}
\centering
\tabcolsep=0.11cm
\begin{tabular}{|c|c|c|c|c|c|}
\hline
ID & $\alpha$ & $\delta$  & $z_{\mathrm{sp}}$ &  $\sigma_{\rm sp}$ [km/s]& $\rm R_{eff}$[kpc]\\
\hline\hline
%A209: & & & & & & & & \\
  GR  &  02:48:03.6 & -03:31:15.7 & 0.194\tnote{a} & $233.1  \pm  12.2$ & $5.03   \pm 0.03$    \\
  BCG &  02:48:03.4 & -03:31:45.0 & 0.189\tnote{a} & $377.8  \pm  15.1$ & $10.45  \pm 0.15$    \\
  G1  &  02:48:02.4 & -03:32:01.9 & 0.191\tnote{a} & $254.9  \pm  12.8$ & $3.35   \pm 0.02$    \\
  G2  &  02:48:03.4 & -03:32:09.3 & 0.195\tnote{a} & $201.8  \pm  15.4$ & $1.59   \pm 0.02$    \\
 15   &  02:48:03.0 & -03:30:18.2 & 0.188\tnote{a} & $141.9  \pm  16.0$ & $2.11   \pm 0.02$    \\
 16   &  02:48:03.0 & -03:30:20.8 & 0.195\tnote{a} & $273.4  \pm  13.5$ & $2.35   \pm 0.01$    \\
 146  &  02:48:00.5 & -03:31:21.6 & 0.191\tnote{a} & $121.5  \pm  34.5$ & $1.25   \pm 0.01$    \\
 223  &  02:48:02.1 & -03:30:43.9 & 0.194\tnote{a} & $194.8  \pm  11.1$ & $3.67   \pm 0.02$    \\
 410  &  02:48:03.7 & -03:31:02.0 & 0.182\tnote{a} & $159.6  \pm  24.8$ & $1.73   \pm 0.01$\\
 658  &  02:48:08.5 & -03:31:28.9 & 0.195\tnote{a} & $207.7  \pm  13.4$ & $2.49   \pm 0.01$    \\
 683  &  02:48:00.3 & -03:31:29.2 & 0.179\tnote{a} & $164.3  \pm  33.1$ & $0.62   \pm 0.01$    \\
 711  &  02:48:05.9 & -03:31:31.9 & 0.186\tnote{a} & $159.4  \pm  13.7$ & $1.78   \pm 0.01$    \\
 770  &  02:48:03.7 & -03:31:35.0 & 0.190\tnote{a} & $172.1  \pm  18.3$ & $1.16   \pm 0.02$      \\
 773  &  02:48:02.8 & -03:31:47.1 & 0.186\tnote{a} & $212.0  \pm  12.3$ & $3.16   \pm 0.03$    \\
 816  &  02:48:08.3 & -03:31:39.2 & 0.191\tnote{a} & $192.8  \pm  16.2$ & $4.21   \pm 0.05$    \\
 906  &  02:48:03.7 & -03:31:58.4 & 0.190\tnote{a}  & $240.4  \pm  20.5$ & $1.74   \pm 0.02$ \\
 975  &  02:48:01.0 & -03:31:54.7 & 0.192\tnote{a} & $75.0   \pm  35.4$ & $0.97   \pm 0.01$      \\
 1034 &  02:48:07.1 & -03:31:46.9 & 0.184\tnote{a} & $81.0   \pm  48.1$ & $2.22   \pm 0.01$      \\
 1069 &  02:48:04.5 & -03:32:06.5 & 0.196\tnote{a} & $244.6  \pm  16.0$ & $3.25   \pm 0.01$      \\
 1214 &  02:48:05.4 & -03:32:18.4 & 0.185\tnote{a} & $212.0  \pm  21.8$ & $2.0    \pm 0.01$      \\
% 1235 &  02:48:03.2 & -03:32:18.6 & 0.194\tnote{a} & $33.5    \pm 40.9$ & $1.99    \pm0.01$      \\
 1479 &  02:48:04.9 & -03:32:36.7 & 0.183\tnote{a} & $108.5  \pm  20.3$ & $1.03   \pm 0.01$      \\
 208   &  02:48:02.6 & -03:30:37.7 & 0.184\tnote{b} &    -  &    -      \\
 233   &  02:48:03.7 & -03:30:43.2 & 0.186\tnote{b} &    -  &    -      \\
 367   &  02:48:06.8 & -03:30:55.3 & 0.197\tnote{b} &    -  &    -      \\
 496   &  02:48:05.1 & -03:31:10.1 & 0.191\tnote{b} &    -  &    -      \\
 601   &  02:48:04.5 & -03:31:19.6 & 0.193\tnote{b} &    -  &    -      \\
 742   &  02:48:02.8 & -03:31:32.8 & 0.188\tnote{b} &    -  &    -      \\
 792   &  02:48:04.6 & -03:31:34.5 & 0.184\tnote{b} &    -  &    -      \\
 901   &  02:48:01.9 & -03:31:45.8 & 0.203\tnote{b} &    -  &    -      \\
 1274  &  02:48:01.1 & -03:32:23.3 & 0.197\tnote{b} &    -  &    -      \\
 1342  &  02:48:00.6 & -03:32:26.8 & 0.188\tnote{b} &    -  &    -      \\
 1362  &  02:48:05.5 & -03:32:30.4 & 0.192\tnote{b} &    -  &    -      \\
% 1522  &  02:48:03.8 & -03:32:40.2 & 0.187\tnote{b} &    -  &    -      \\
 1551  &  02:48:05.3 & -03:32:44.0 & 0.188\tnote{b} &    -  &    -      \\
 1670  &  02:48:05.9 & -03:32:53.0 & 0.191\tnote{b} &    -  &    -      \\
\hline\hline
\end{tabular}\footnotesize
\begin{tablenotes}
       \item[a] From the Hectospec Survey
       \item[b] From the VIMOS CLASH-VLT Survey
     \end{tablenotes}%\end{centre}
     \end{threeparttable}
\end{table}

\section{Cluster members}
\label{sec:catalogs}
\begin{figure}
\centering
\includegraphics[width=8.5cm]{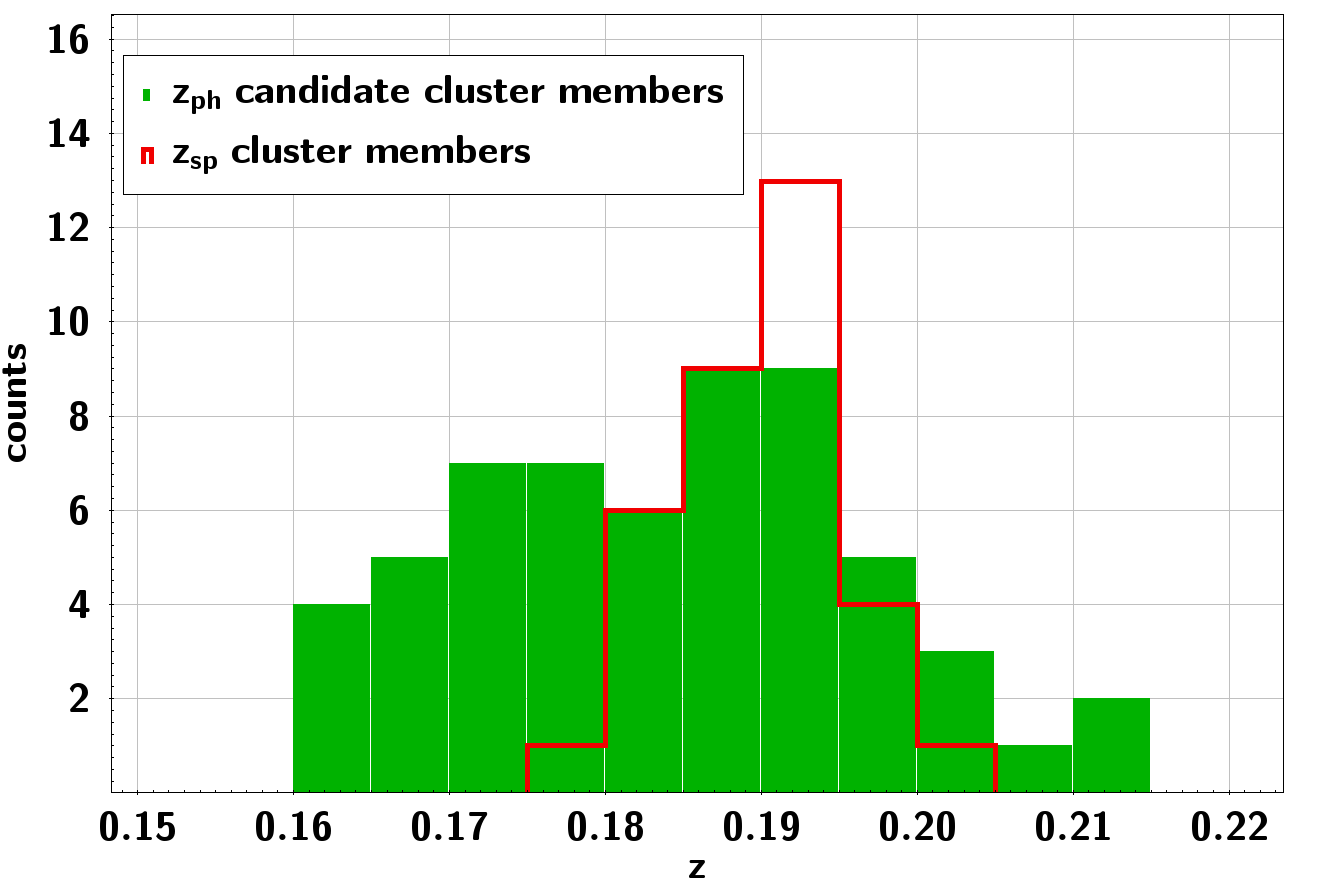}
\caption{\small Redshift distribution of the final cluster member sample, including photometric candidates (green histogram) and spectroscopically confirmed members (red histogram).}
\label{fig:z_cl_histo}
\end{figure}

\begin{figure}
\centering
\includegraphics[width=8.7cm]{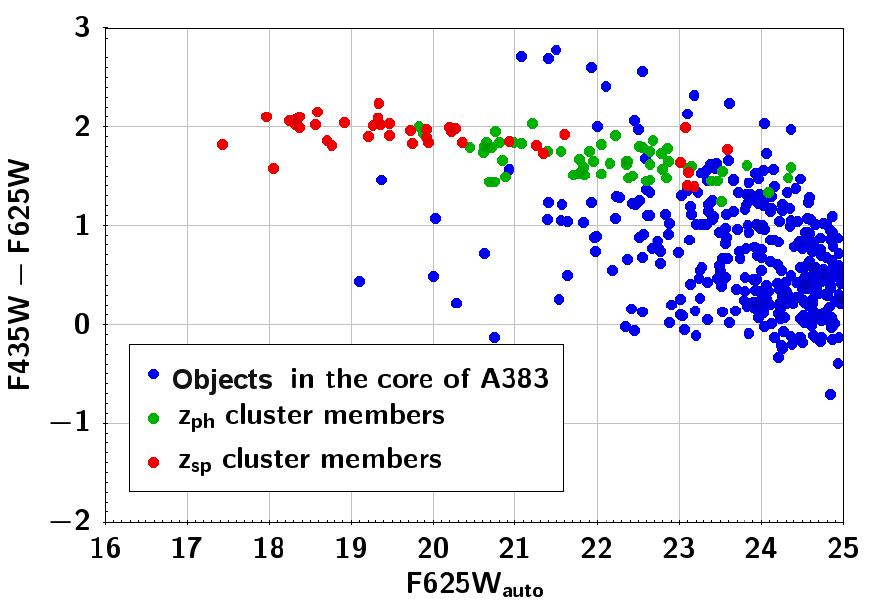}
\caption{\small Colour magnitude diagram for the sources extracted in the  core of A383.
We plot the colour from aperture magnitudes in the filters F435W and F625W versus the \texttt{SExtractor} mag$\_$auto in the F625W filter.
Blue circles are all the sources extracted in the cluster core; in red we plot the spectroscopically confirmed cluster members and in green the photometric cluster member candidates with $z\rm_{ph}\in[0.16,0.22]$.}  
        \label{fig:col_mag}
\end{figure}

In order to define the galaxy component to include in the strong lensing analysis (see Section~\ref{sec:lensing}), we select cluster members in the core of A383 combining  the photometric and spectroscopic datasets.
We restrict our analysis to the sources in a FOV of $1.5'\times1.5'$ centred on the cluster.
In this FOV we have 34 spectroscopically confirmed cluster members (13 from VLT/VIMOS data and 21 from Hectospec data), which have $|z_{\mathrm{sp}}-z_{\mathrm{cl}}|<0.01$ (see Fig.~\ref{fig:a383rgb}),  where $z_{\mathrm{cl}}=0.189$ is the cluster redshift. 
To include  in our lensing analysis also cluster members which lack spectroscopic data, we select further member candidates combining information from the cluster colour-magnitude diagram and from photometric redshifts.  
We  compute photometric redshifts for the galaxies extracted in our dataset using the spectral energy distribution (SED) fitting code \texttt{LePhare}\footnote{http://www.cfht.hawaii.edu/~arnouts/lephare.html} \citep{Arnouts1999,Ilbert2006}.
We use  the COSMOS template set \citep{Ilbert2009} as galaxy templates, including  31 galaxy SEDs for elliptical, spiral and starburst galaxies.
We apply the Calzetti extinction law \citep{Calzetti2000} to the starburst templates, and the SMC Prevot law \citep{Prevot1984} to the Sc and Sd templates, to take into account extinction due to the interstellar medium (ISM).
As we did in \citet{Monna2014}, in order to account for template mismatch of red SEDs \citep{Greisel2013}, we apply offset corrections to our photometry. 
We use the sample of spectroscopically confirmed cluster members to estimate these photometric offsets through a colour adaptive method. 
For each galaxy with  known $z_{\mathrm{sp}}$, the tool finds the template which best fits its observed photometry, and thus minimises the offset between the observed and predicted magnitudes in each filter.  
\\The photometric redshift and spectroscopic measurements for the relatively bright spectroscopically confirmed cluster members present small scatter: all of them fall within $|z_{\mathrm{sp}}-z_{\mathrm{ph}}|<0.02$.  
However, we use a larger interval of  $|z_{\mathrm{ph}}-z_{\mathrm{cl}}|<0.03$ to select candidate cluster members photometrically, since faint galaxies have larger photometric redshift errors in general.
In addition, we require candidate cluster members to be brighter than 25 mag in the F625W filter ($ \rm F625W_{auto}<25$) and to lie around the red sequence in the colour$-$magnitude diagram (having $\rm F435W-F625W\in[1.3,2.3]$, see Fig.~\ref{fig:z_cl_histo},~\ref{fig:col_mag}). 
\\ 

Our final cluster member sample in the core of the cluster contains 92 galaxies, 34 spectroscopically confirmed and 58 photometric candidates.

\section{Strong Lensing recipe}
\label{sec:lensing}
We perform the strong lensing analysis of A383  using the strong lensing parametric mass modelling 
software \texttt{GLEE} \citep{Suyu2010,Suyu2012}. 
As constraints we use the positions and redshifts (when known) of the multiple images.
These directly measure the differences of  lensing deflection angles at the position of multiple images. 
In addition, we also reconstruct the surface brightness distributions of giant arcs, which contain information on higher order derivatives of the deflection angle. 
We adopt analytic mass models to describe the mass profiles of the cluster dark halo and the galaxy mass components. 
The best fitting model is found through a simulated annealing minimisation in the image plane. 
The most probable parameters and uncertainties for the cluster mass model are then obtained from  a Monte Carlo Markov Chain (MCMC) sampling. 
 
\subsection{Mass components and scaling relations}
We describe the smooth dark halo (DH) mass component of the cluster with a Pseudo Isothermal Elliptical Mass Distribution (PIEMD) profile \citep{Kassiola1993}.
Its projected  surface density is
 \begin{equation}
 \Sigma(R)= \frac{\sigma^2}{2 {\rm G}}\left(\frac{1}{\sqrt{r_{\rm c}^2+R^2}}\right),
 \end{equation}
where $\sigma$ is the velocity dispersion of the DH and $r\rm_c$ is its core radius. 
$R$ is the 2D radius, given by $R^2=x^2/(1+e)^2+y^2/(1-e)^2$ for an elliptical profile with ellipticity $e$. 
Strong lensing allows to robustly  constrain the Einstein radius of a lens. This is the radius of the Einstein ring formed by a point source  lensed by a spherical halo, when source and lens are aligned with the observer. For a singular isothermal sphere the Einstein radius $\theta_E$ and velocity dispersion of the halo are related by
\begin{equation}
\theta_E=4\pi\left(\frac{\sigma}{c}\right)^2\frac{D_{ds}}{D_s}=\Theta_E\frac{D_{ds}}{D_s}
\end{equation}
where $\theta_E$ and $\Theta_E$ are in arcseconds,  $c$ is the speed of light, $D_s$ is the distance of the lensed source and $D_{ds}$ is the distance between the lens and the source. 
$\Theta_E$ is the \textit{Einstein parameter}, which corresponds to the Einstein radius for $D_{ds}/D_s=1$. 
For elliptical mass distribution with core radius, lensing measures the Einstein parameter $\Theta_E$, which corresponds to the Einstein radius when the ellipticity and core radius go to zero $(e, r_{c}\rightarrow0)$. 
In the following analysis we use the Einstein parameter $\Theta_E$ to describe the mass amplitude of the lens halo.\\
The total mass associated with each cluster member is modelled with a dual pseudo isothermal elliptical profile (dPIE) \citep{Elisa2007}. 
This model has a core radius $r_c$ and a truncation radius $r_{\rm tr}$, which marks the region where the density slope changes from $\rho\propto r^{-2}$ to $\rho\propto r^{-4}$.\\
 The projected surface mass density is
 \begin{equation}
 \Sigma(R)= \frac{\sigma^2}{2 {\rm G} R}\frac{r_{\rm tr}^2}{(r_{\rm tr}^2-r_c^2)} \left(\frac{1}{\sqrt{1+\frac{r_c^2}{R^2}}}-\frac{1}{\sqrt{1+ \frac{r_{\rm tr}^2}{R^2}}}\right)
 \label{eq:bbs}
 \end{equation}
 where $R^2=x^2/(1+e)^2+y^2/(1-e)^2$, as for the PIEMD mass profile.
The total mass is given by
\begin{equation}
  M_{\rm tot}=\frac{\pi\sigma^2}{G}\frac{r_{\rm tr}^2}{r_{\rm tr}+r_c}\,\,,
  \label{eq:m_bbs1}
\end{equation}
which, for $r_c\rightarrow0$, reduces to
\begin{equation}
 M_{\rm tot}=\frac{\pi\sigma^2r_{\rm tr}}{G}\,\,.
 \label{eq:m_bbs}
\end{equation}
For vanishing core radius, $r_{\rm tr}$ corresponds to the radius containing half of the total mass of the galaxy \citep[see Appendix A3 in][]{Elisa2007}. 
%Thus $r_{\rm tr}$ can be considered as a half mass radius.   
We adopt vanishing core radii for the galaxies (unless stated otherwise), so that we have only 2 free parameters associated with each galaxy, i.e. $\sigma$ and $r_{\rm tr}$.
However, we selected 92 cluster members in the cluster core and this would yield $\sim200$ free parameters for the galaxy mass component. 
To reduce this large number of free parameters, we adopt luminosity scaling relations to relate the velocity dispersion and truncation radius of the cluster members to a fiducial reference galaxy, as in \cite{Halkola2007} and \cite{Eichner2013}. 
In other words, we only optimize $\sigma$ and $r_{\rm tr}$ for a  reference galaxy and then scale all the other galaxies' $\sigma$ and $r_{\rm tr}$ through luminosity scaling relations (i.e. the Faber-Jackson and fundamental plane for $\sigma$ and $r_{\rm tr}$, respectively). \\
Given the Faber-Jackson relation, the central velocity dispersion of early type galaxies is proportional to a power law of the luminosity. 
Thus for the cluster member we adopt 
\begin{equation}
\sigma=\sigma_{GR}\left(\frac{L}{L_{GR}}\right)^{\delta}  
\label{eq:F_J}
\end{equation}
where the amplitude $\sigma_{GR}$ is the velocity dispersion of a reference galaxy halo with luminosity
$L_{GR}$. \\
Following \citet{Hoekstra2003}, \citet{Halkola2006, Halkola2007} and \citet{Limousin2007}, we assume that the truncation radius of galaxy halos scales with luminosity as
\begin{equation}
r_{\rm tr}=r_{\rm tr,GR}\left(\frac{L}{L_{GR}}\right)^\alpha=r_{\rm tr,GR}\left(\frac{\sigma}{\sigma_{GR}}\right)^\frac{\alpha}{\delta}  
\label{eq:r_tr}
\end{equation}
where $r_{\rm tr,GR}$ is the truncation radius for a galaxy with luminosity $L_{GR}$. 
Given  Eq.\,\ref{eq:F_J} and Eq.\,\ref{eq:r_tr}, once we fix the exponent $\delta$ and $\alpha$, the free parameters, used to tune the galaxy mass contribution to the total cluster mass, are reduced to the velocity dispersion $\sigma_{\rm GR}$ and truncation radius $r_{\rm tr,GR}$ of the reference galaxy.\\
The total mass-to-light ratio for a galaxy scales as
\begin{equation}
 \frac{M_{\rm tot}}{L}\propto L^\epsilon\propto\sigma^{\frac{\epsilon}{\delta}}
 \label{eq:ML}
\end{equation}
and it is constant for $\epsilon=0$. Combining Eq. \ref{eq:m_bbs}, \ref{eq:F_J} and \ref{eq:r_tr}, the total mass scales as
\begin{equation}
 M_{\rm tot}\propto\sigma^2 r_{\rm tr}\propto \sigma^{2+\frac{\alpha}{\delta}}\rm .
 \label{eq:m1}
 \end{equation}
Therefore, from Eq.\,\ref{eq:ML} and Eq.\,\ref{eq:m1}, we obtain the following relation for the exponents
\begin{equation}
 \alpha=\epsilon -2\delta+1\,,
 \label{eq:m2}
\end{equation}
 which means that if we have knowledge of two of them, we can derive the third one.
 In the following we will go through some considerations which will help us to fix the value of these exponents.\\
For elliptical galaxies in clusters, the exponent $\delta$ has measurements between 0.25 and 0.3, depending on the filter in which the photometry is extracted \citep[see][]{Ziegler1997,Fritz2009,Kormendy2013,Focardi2012}. 
Measurements from strong and weak lensing analyses yield $\delta=0.3$ \citep[see][]{Rusin2003, Brimioulle2013}.  \\
Concerning the truncation radius of galaxies in clusters,
%we expect that it scales with the total mass as $ \rm r_{\rm tr}\propto M_{\rm tot}^\frac{1}{3}$ 
theoretical studies predict that in dense environment it scales linearly with the galaxies velocity dispersion 
\citep[see][]{Merritt1983} and, given Eq.\,\ref{eq:m_bbs}, this yields $M_{\rm tot}\propto\sigma^3$. 
However, from Eq.\,\ref{eq:ML}, the total galaxy mass can be written as $M_{\rm tot}\propto\sigma^{\frac{\epsilon+1}{\delta}}$, and thus we conclude that $\epsilon=3\delta-1$. \\
In summary, we expect $\delta$ to be within $[0.25,0.3]$, which implies $\epsilon$ to be within $[-0.3,0.2]$.
The lower limit $\epsilon=-0.3$ is the case in which the galaxy halos mass has a value as expected 
for completed stripping process \citep[e.g., see][]{Merritt1983}. 
%is the case in which the dark halo of the galaxy in the cluster has undergone a complete stripping to the expected value, due to the interaction with the cluster dark halo and the other cluster members.  
On the contrary, $\epsilon=0.2$ is the case in which the galaxies have suffered no stripping at all and fullfill the scaling relations in fields \citep[e.g., see][]{Brimioulle2013}. \\
A383 is a relaxed galaxy cluster, thus we expect the galaxy halo stripping process to be completed in the core. 
However, to take into account still ongoing halo stripping, as we did in \cite{Eichner2013} we fix the exponents of the mass to light luminosity relations to be $\epsilon=0$. 
This value is in between the ones expected for not yet started, and already completed halo stripping.\\ 
Using the sample of confirmed cluster members with measured velocity dispersions, we directly measured the exponent $\delta$ of the Faber-Jackson relation in the F814W band. 
We get $\delta\sim0.296$, thus in the lensing analysis we use $\delta=0.3$. 
%In \cite{Eichner2013} we  confirmed that using $ \delta=0.3$ or $ \delta=0.25$ does not affect the results on the halos sizes. 
\\
Finally, referring to the general relation between the exponents of the scaling relations given in Eq.\ref{eq:m2}, we obtain that $\alpha=0.4$.\\
Once we fix the exponents of the luminosity scaling relations, the only parameters we need to determine to define the galaxy mass component are the amplitudes of the scaling relations, $\sigma_{\rm GR}$ and $r_{\rm tr,GR}$. 
We use as reference galaxy (GR) the third brightest galaxy of the cluster \mbox{(RAJ2000=02:48:03.63,}  \mbox{DECJ2000=-03:31:15.7)}, which has $\rm F814W\_iso=17.74\pm0.01$, $z_{\mathrm{sp}}=0.194$ and a measured velocity dispersion $\sigma_{\rm GR}=233\pm12$\,km/s.\\
Bright galaxies, as the cluster BCGs, can show a large scatter and likely also a systematic deviation from the scaling relations of cluster luminous red galaxies \citep[see][]{Postman2012b,Kormendy2013}.
Thus we model the BCG independently to better account for its contribution to the total mass profile. 
Moreover we independently optimise two further cluster members close to the lensed system 3-4 (see Sec.\,\ref{sec:extended_image}). 
These galaxies have measured $\sigma_{\rm sp}$, which, combined with the strong lensing constraints from their nearby arcs, allow to directly measure their halo sizes.
We call these two galaxies  G1 and G2 (see Fig.~\ref{fig:a383rgb}). Their redshifts and measured velocity dispersions are $ z_{\rm G1}=0.191$, $z_{\rm G2}=0.195$ and  $\sigma_{\rm G1}=255\pm13$\,km/s, $\sigma_{\rm G2}=202\pm15$\,km/s, respectively (see Tab.\,\ref{tab:vel_gal}).\\
To scale $\sigma$ and $r_{\rm tr}$ of the cluster members according to Eq.~\ref{eq:F_J} and ~\ref{eq:r_tr}, we use the observed isophotal fluxes in the F814W filter. 
Moreover, assuming that the luminosity of the galaxies traces their dark matter halos, we fix the ellipticity and orientation of each halos to the respective values associated with the galaxy light profile, as extracted with \texttt{SExtractor} in the F814W band.\\
In addition, we also allow for an external shear component to take into account the  large scale environment contribution to the lensing potential.

\subsection{Multiple images}
We use the 9 systems of multiply lensed sources presented in \citet{Zitrin2011} as constraints for our lens modelling. 
Four of these systems (system 1 to 4 in Tab.~\ref{tab:multiple images})  are well known and spectroscopically confirmed and were used in previous lensing analyses \citep{Newman2011,Sand2004,Sand2008,Smith2005}. 
System 5 is a double lensed  source at z=6, which has been spectroscopically confirmed by \citet{Richard2011}. 
System 6, identified by \citet{Zitrin2011}, has been followed up with VIMOS in the spectroscopic CLASH-VLT survey and confirmed to be at $z_{\rm s}=1.83$. 
Systems 7 to 9 lack spectroscopic data, thus we estimate their photometric redshifts with \texttt{LePhare}.
For these systems we adopt as source redshift $\rm z_{SL}$ the photometric redshift of the brightest multiple images with photometry uncontaminated by nearby galaxies. 
Altogether we have 27 multiple images of 9 background sources, of which 6 are spectroscopically confirmed lensed sources (systems 1-6).
In Tab.~\ref{tab:multiple images} we list the positions and redshifts for all the images. \\

For the systems with spectroscopic confirmation,  we fix the source redshift $\rm z_{SL}$ in the lens model to the spectroscopic value $\rm z_{sp}$. 
For the other systems the $\rm z_{SL}$ are free parameters. Their photometric predictions are used as starting values for $\rm z_{SL}$ and we optimise them with  gaussian priors. 
As  widths of the gaussian priors we adopt 3 times the uncertainties of the photometric redshifts.
This is to explore a range of source redshifts $z_{\rm SL}$  larger than the range indicated by the $1\sigma$ uncertainties of the $z_{\rm ph}$.  \\
Using the HST photometric dataset we can estimate the positions of multiple images with a precision of $0.065\arcsec$.
\citet{Host2012} and \citet{D'Aloisio} estimated that, on cluster scales, multiple image positions are usually reproduced with an accuracy of $\sim$1-2 arcseconds due to structures along the line of sight. 
\citet{Grillo2014} show that a higher precision can be reached through a detailed strong lensing analysis of the cluster core. They predict the positions of  the observed multiple images in the core of MACS J0416 with a median offset of 0.3\arcsec.
In this work, we adopt errors of $1\arcsec$ on the position of the observed multiple images to account for uncertainties due to density fluctuations along the line of sight.
\begin{table}
\caption{Summary of the multiply lensed systems used to constrain the strong lensing model of A383 (see also \citet{Zitrin2011}).
  The columns are: Col.1 is the ID; Col.2-3 Ra and Dec; Col.4  is the source redshift $z_s$, for systems 1-6 it is the spectroscopic redshift $z_{\rm sp}$ from VLT/Vimos (see text), for systems 7-9 we give the photometric redshift $z_{\rm ph}$  with the $3\sigma$  uncertainties estimated for the multiple image with the best photometry; Col.5 gives the strong lensing predictions for the sources redshifts from the  model performed including the measured velocity dispersions, with the respective $1\sigma$ uncertainties.}
\footnotesize
\begin{tabular}{|l|c|c|c|c|}
\hline
\hline
 Id & Ra & Dec & $\rm z_s$ & $\rm z_{SL}$\\
 \hline 
 1.1       &  02:48:02.33& -03:31:49.7 &  1.01 &  1.01 \\
 1.2      &  02:48:03.52& -03:31:41.8 &   "   &   "   \\
 \hline     
 2.1       &  02:48:02.95& -03:31:58.9 &  1.01 &  1.01 \\
 2.2       &  02:48:02.85& -03:31:58.0 &    "  &    "  \\      
 2.3       &  02:48:02.45& -03:31:52.8 &    "  &    "  \\
 \hline     
 3.1       &  02:48:02.43 & -03:31:59.4 &  2.58 &  2.58   \\
 3.2       &  02:48:02.31 & -03:31:59.2 &  "    &  "    \\ 
 3.3       &  02:48:03.03 & -03:32:06.7 &  "    &  "     \\
 3.4       &  02:48:02.30 & -03:32:01.7 &  "    &  "      \\
 \hline     
 4.1       & 02:48:02.24 &-03:32:02.1  &  2.58 &  2.58   \\
 4.2       & 02:48:02.21 &-03:32:00.2  &  "    &  "      \\
 4.3       & 02:48:02.85 &-03:32:06.7  &  "    &  "      \\
 \hline     
 5.1        &02:48:03.26 &-03:31:34.8  &  6.03 &  6.03  \\
 5.2        &02:48:04.60 &-03:31:58.5  &   "   &   "     \\
 \hline     
 6.1       &02:48:04.27 &-03:31:52.8   &  1.83&  1.83\\ 
 6.2       &02:48:03.38 &-03:31:59.3   &   "  &   "  \\ 
 6.3       &02:48:02.15 &-03:31:40.9   &  "   &  "   \\ 
 6.4       &02:48:03.72 &-03:31:35.9   &  "   &  "   \\  
 \hline     
 7.1       &02:48:04.09  &-03:31:25.5 & 4.46   [3.71,5.09] &  $4.94^{+0.30}_{-0.28}$\\
 7.2       &02:48:03.57  &-03:31:22.5  & "& "        \\
 7.3       &02:48:03.13  &-03:31:22.2  & "&  "\\
 \hline                    
 8.1      &  02:48:03.68&-03:31:24.4   &   2.3   [1.85,3.38]& $1.78^{+0.31}_{-0.23}$ \\
 8.2      &  02:48:03.39 &-03:31:23.5 &   "&" \\
 \hline    
 9.1      & 02:48:03.92&-03:32:00.8    & " & $4.10^{+0.56}_{-0.68}$\\
 9.2      & 02:48:04.05& -03:31:59.2   &" &"\\
 9.3     & 02:48:03.87&-03:31:35.0   &  " &"\\
 9.4      & 02:48:01.92& -03:31:40.2  &  3.45 [3.30,3.60]&"\\
                                
\hline     
 \hline 
  \end{tabular}\\
\label{tab:multiple images}
\end{table}
\section{Pointlike models}
\label{sec:pointlike_model}
We now carry on with the strong lensing modelling of A383, using as constraints the observed positions and spectroscopic redshifts of the multiple images listed in Table \ref{tab:multiple images}. \\   
In this Section we investigate  how much the precision of the lens model and the constraints on the $r_{\rm tr}$ of the galaxies improve when we use the velocity dispersion measurements as inputs for the lens model.  
Thus we construct two parallel models.\\
In the first model we scale all the galaxies with respect to GR using Eq.~\ref{eq:F_J} and the left side of Eq.~\ref{eq:r_tr}. We individually optimise only the BCG, the reference galaxy GR and the two galaxies G1 and G2 close to the lensed system 3-4. Their velocity dispersions and truncation radii are optimised with flat priors in the range of [100,500]\,km/s and [1,100]\,kpc.\\
In the second model we fix the velocity dispersions of  the 21 cluster members from the Hectospec survey to their measured values. Their truncation radii are then given by the left side of Eq.\,\ref{eq:r_tr}. All the other cluster members are still scaled with respect to GR according to Eq.\,\ref{eq:F_J} and Eq.\,\ref{eq:r_tr}.
In the lens modelling we allow for some freedom for the velocity dispersions of the BCG, GR, G1 and G2, which we optimise around their $\sigma_{\rm sp}$  using gaussian priors with width equal to their spectroscopic uncertainties. Also in this case the truncation radii of these four galaxies are optimised with a flat prior within [1,100]kpc.\\
We will refer to these two models as ``pointlike models with and without velocity dispersions'' (hereafter ``$\rm w/\sigma$'' and ``$\rm wo/\sigma$'' respectively). With ``pointlike'' we  indicate that multiple image constraints are used as points, without accounting, at this stage, for surface brightness constraints.\\
\begin{figure*}
 \centering
 \includegraphics[width=18cm]{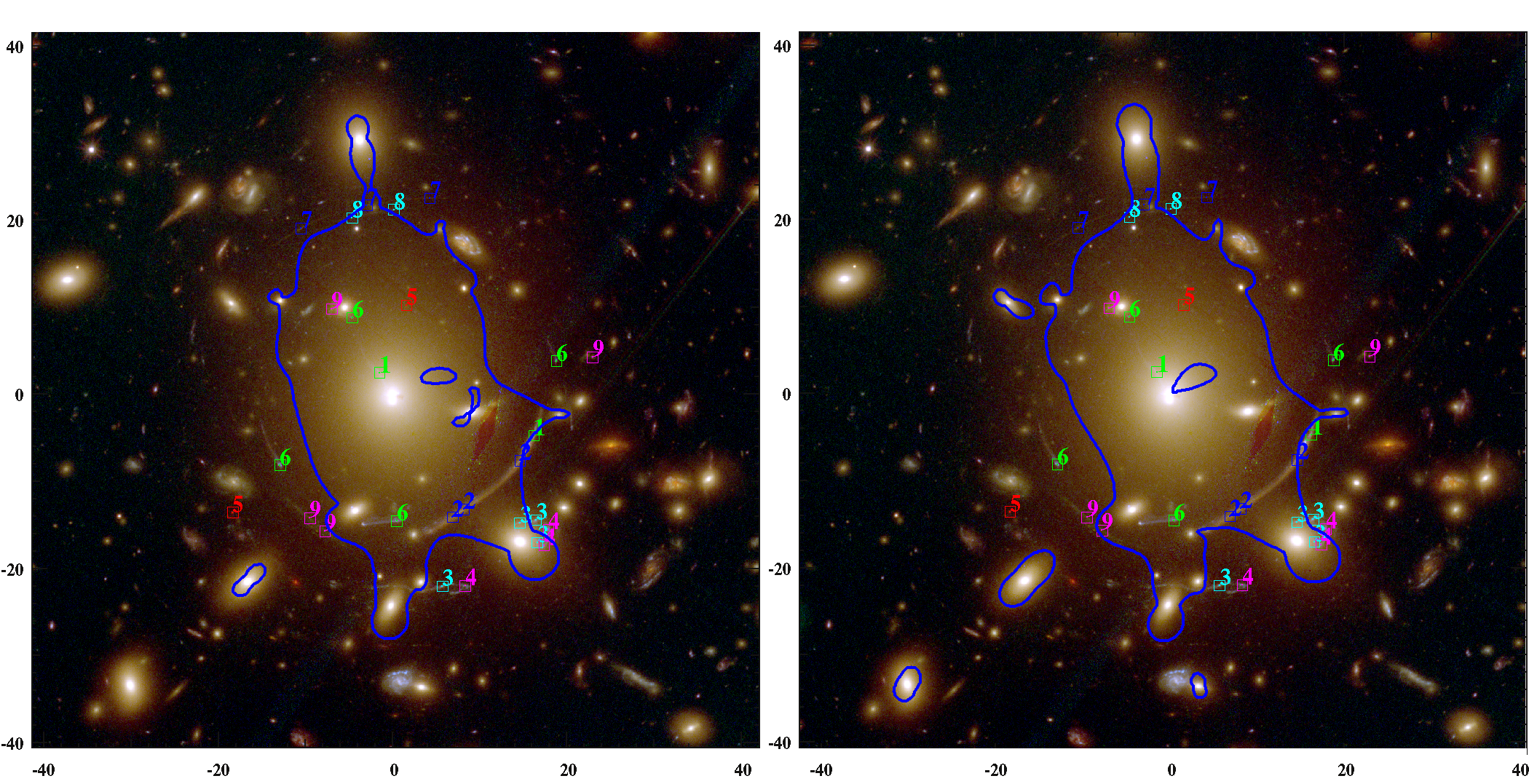}
 \caption{\small HST colour composite image of A383 core with the critical lines overplotted (in blue) for the SL models. Left panel shows the model ``$\rm wo/\sigma$'', and the right panel is the model including the measured velocity dispersions. The critical lines are for a source at  $z_s=2.58$, which is the spectroscopic redshift of system 3-4.}
         \label{fig:a383_SL}
 \end{figure*}
 \begin{figure}
\centering
\includegraphics[width=9cm]{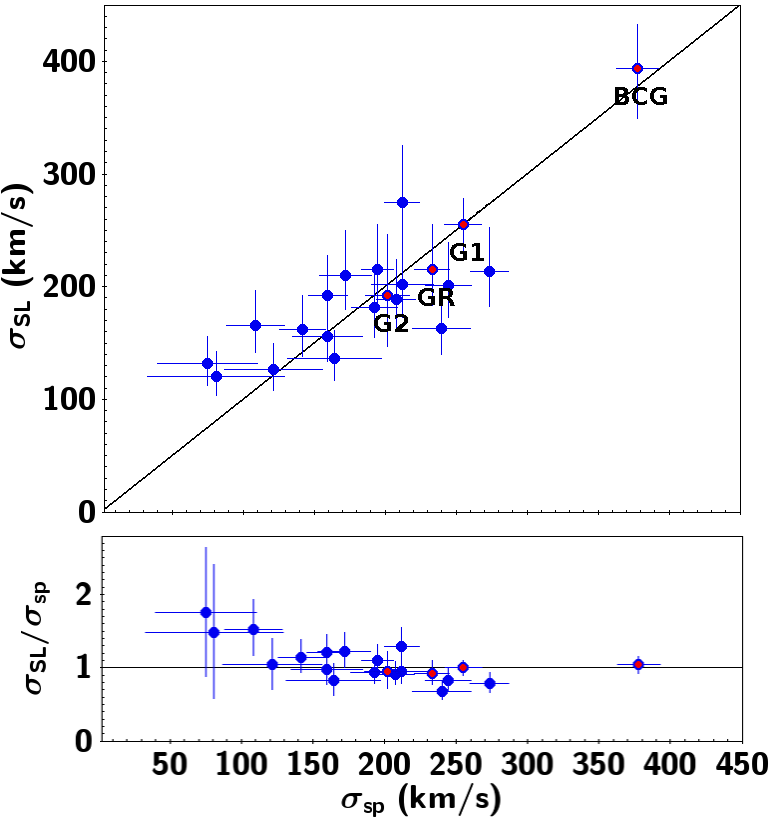}
\caption{\small Upper panel: velocity dispersions $\sigma_{\rm SL}$ predicted from the SL analysis versus measured velocity dispersions $\sigma_{\rm sp}$  for the 21 cluster members from the Hectospec survey. Lower panel: ratio of  $\sigma_{\rm SL}$ and $\sigma_{\rm sp}$. The $\sigma_{\rm SL}$ are predicted using the scaling luminosity relation, except for the four galaxies individually optimised, GR, BCG, G1 and G2. The values predicted in the model ``$\rm wo/\sigma$'' are globally in agreement with the measured $\sigma_{\rm sp}$ at the $1\sigma$ level, except at low velocity dispersions where they are slightly overestimated by a factor of $\sim1.5$. 
We label in red the data for the galaxies individually optimised: these show an excellent agreement between the $\sigma_{\rm SL}$ and the measured $\sigma_{\rm sp}$.}  
        \label{fig:sigma_sl_sp}
\end{figure}
In both cases, we optimise all the DH parameters using flat priors: the DH centre is optimised within 3 arcsec  from the BCG position, ellipticity within [0,1], the position angle (PA) is free to vary  within $180^\circ$ and the core radius within [0,60]\,kpc. The Einstein parameter $\Theta_E$ is optimised within  $[4.5,65]$ arcseconds, which correspond to the  velocity dispersion range [400,1500]\,km/s for a singular isothermal sphere.
%In the surface density 
For each galaxy  we fix its position, ellipticity $\epsilon$ and position angle PA to the value measured from the photometry in the ACS/F814W filter. 
Only for the BCG we optimise the values of $\epsilon$ and PA measured in the ACS/F814W filter  with a gaussian prior with width of 0.25 and $10^\circ$ respectively.
The two models have 18 free parameters associated with the mass components.
In Tab.~\ref{tab:lensing_models} we list the results on the galaxy and DH  parameters for both models and in the appendix (Fig.~\ref{fig:mcmc_dh_sig},\ref{fig:mcmc_gal_no_sig}) we provide the plots of the MCMC sampling.\\
Here we summarise the main results.\\
The final best model $\rm wo/\sigma$ reproduces the positions of the observed multiple images to an accuracy of 0.5\arcsec, with  $\rm\chi^2=0.6$  in the image plane.   
The cluster dark halo has a core radius of $\sim37.5_{-7.7}^{+5.6}$ kpc and Einstein parameter $\Theta_E=13.3_{-2.2}^{+2.6}$ arcsec, which corresponds to the fiducial Einstein radius $\theta_E=11.6_{-1.9}^{+2.3}$ arcsec for a source at $z_{\rm s}=2.58$. It gives a central velocity dispersion of $\rm\sigma= 680_{-57}^{+67}$\,km/s for a singular isothermal sphere.
The BCG has a velocity dispersion of $\rm \sigma_{BCG}=395_{-44}^{+39}$\,km/s and truncation radius of $r_{\rm tr}=53.1_{-25.0}^{+15.6}$ kpc. 
The predicted velocity dispersion and radii for  GR, G1 and G2 are   $\rm \sigma_{GR}=214_{-32}^{+40}$\,km/s, $\rm \sigma_{G1}=253\pm23$\,km/s, $\rm \sigma_{G2}=194_{-45}^{+54}$\,km/s and $r_{\rm tr,GR}=23.1_{-12.8}^{+29.4}$ kpc,  $r_{\rm tr,G1}=47.8\pm20.9$ kpc,  $r_{\rm tr,G2}=32.2_{-23.4}^{+31.5}$ kpc. 
The total mass of the cluster within the Einstein radius $\theta_E$=$11.6_{-1.9}^{+2.3}$ arcsec is $M_{\rm tot}=9.72\pm0.23\times10^{12}M_\odot$. 

Including the measured velocity dispersions in the strong lensing analysis leads to a final best model with $\rm\chi^2=0.5$ in the image plane, which reproduces the multiple images positions with a mean  accuracy of 0.4\arcsec. 
The smooth dark halo has core radius of $39.5_{-5.7}^{+5.3}$\,kpc, and  $\theta_E$=$11.1_{-1.6}^{+2.1}$ arcsec for a source at $z_{\rm s}=2.58$, from which we get $\rm\sigma= 667_{-47}^{+62}$\,km/s for a singular isothermal sphere. 
The measured velocity dispersions of GR, BCG, G1 and G2 are optimised within their uncertainties. The final values for these parameters are $\rm \sigma_{GR}=238\pm15$\,km/s, $\rm \sigma_{BCG}=379\pm21\,$km/s, $\rm \sigma_{G1}=252\pm14$\,km/s and $\rm \sigma_{G2}=201\pm20$\,km/s. 
The predicted radii for  GR, BCG, G1 and G2 are $r_{\rm tr,GR}=13.2_{-4.3}^{+6.2}$ kpc, $r_{\rm tr,BCG}=58.4_{-33.2}^{+24.9}$ kpc,  $r_{\rm tr,G1}=73.1_{-35.5}^{+38.7}$ kpc,  $r_{\rm tr,G2}=53.2_{-36.3}^{+49.2}$ kpc.
The total mass of the cluster  is $M_{\rm tot}=9.70\pm0.22\times10^{12}M_\odot$ within the Einstein radius $\theta_E$=$11.1_{-1.6}^{+2.1}$ arcsec for a source at $z_{\rm s}=2.58$.\\
The results for the two pointlike models are globally in agreement within their $1\sigma$ errors. 
\begin{table}
\caption{Most probable mass profiles parameters with the respective $1\sigma$ uncertainties for the smooth dark halo, the BCG, GR, G1 and G2 from strong lensing models of A383. In column (1) we give the results for the model without measured velocity dispersions and in column (2) for the model with velocity dispersions.}
\centering
\footnotesize
\begin{tabular}{|c|c|c|}
\hline
\hline
Param& ``$\rm wo/\sigma$'' & ``$\rm w/\sigma$''\\
\hline
\hline
External shear&&\\
\hline
$\gamma$&$0.07\pm0.03$&$0.04\pm0.02$\\
$\theta [^{\circ}]$&$51^{+17}_{-11}$&$37^{+17}_{-26}$\\
\hline
Dark Halo &&\\
\hline
$\delta x$ [\arcsec] &   $0.7\pm0.5$& $1.0\pm0.4$\\
$\delta y$ [\arcsec]&    $1.0\pm0.8$& $2.4_{-0.6}^{+0.4}$  \\
PA $[^{\circ}]$&   $88^{+10}_{-14}$ & $111\pm20$ \\
b/a &  $0.8\pm0.1$ & $0.90\pm0.06$  \\
$\theta_{\rm E}$ [\arcsec]& $11.6_{-1.9}^{+2.3}$&$11.1_{-1.6}^{+2.1}$ \\    
$r_{c}$ [kpc] &   $37.5_{-7.7}^{+5.6}$ & $39.5_{-5.7}^{+5.3}$   \\  
\hline
BCG &&\\
\hline
PA $[^{\circ}]$&   $94^{\circ}\pm23^{\circ}$  &   $98^{\circ}\pm9^{\circ}$      \\
b/a & $ 0.61_{-0.15}^{+0.18}$& $0.60_{-0.13}^{+0.17}$      \\
$\sigma$ [km/s]& $395_{-44}^{+39}$ &$379\pm21$       \\
$r_{\rm tr}$ [kpc] &  $53.1_{-25.0}^{+15.6}$ & $58.4_{-33.2}^{+24.9}$    \\
\hline
GR &&\\
\hline
$\sigma$ [km/s]&$ 214_{-32}^{+40}$ & $238\pm15$        \\
$r_{\rm tr}$ [kpc] & $23.1_{-12.8}^{+29.4}$ & $13.2_{-4.3}^{+6.2}$       \\
\hline
G1 &&\\
\hline
$\sigma$ [km/s]& $253\pm23$ & $252\pm14$    \\
$r_{\rm tr}$ [kpc] &  $47.8\pm20.9$ & $73.1_{-35.5}^{+38.7}$      \\
\hline
G2 &&\\
\hline
$\sigma$ [km/s] & $194_{-45}^{+54}$& $201\pm20$    \\
$r_{\rm tr}$ [kpc] & $32.2_{-23.4}^{+31.5}$&$53.2_{-36.3}^{+49.2}$     \\
\hline                 
\end{tabular}
\label{tab:lensing_models}
\end{table}
 In Fig.~\ref{fig:a383_SL} we show the critical lines for a source at $z=2.58$ for both the models overplotted on the colour composite image of the cluster core. The global models are in agreement, however using the spectroscopically measured velocity dispersions of cluster members  locally affects the mass distribution reconstruction.
In Fig.\,\ref{fig:sigma_sl_sp} we plot the measured velocity dispersions  $\sigma_{\rm sp}$ versus the predicted ones from SL for the model ``$\rm wo/\sigma$'', and in the lower panel their ratio. 
They are in overall agreement within the $1\sigma$ uncertainties. Only few galaxies present a larger deviation from the measured velocity dispersions, but they are anyhow consistent at the $2\sigma$ level. These are faint galaxies ($\rm F814W\_auto\_mag >18.6$) which have spectroscopic velocity dispersion with uncertainties of $\gtrsim30\%$.
\begin{figure}
\centering
\includegraphics[width=9cm]{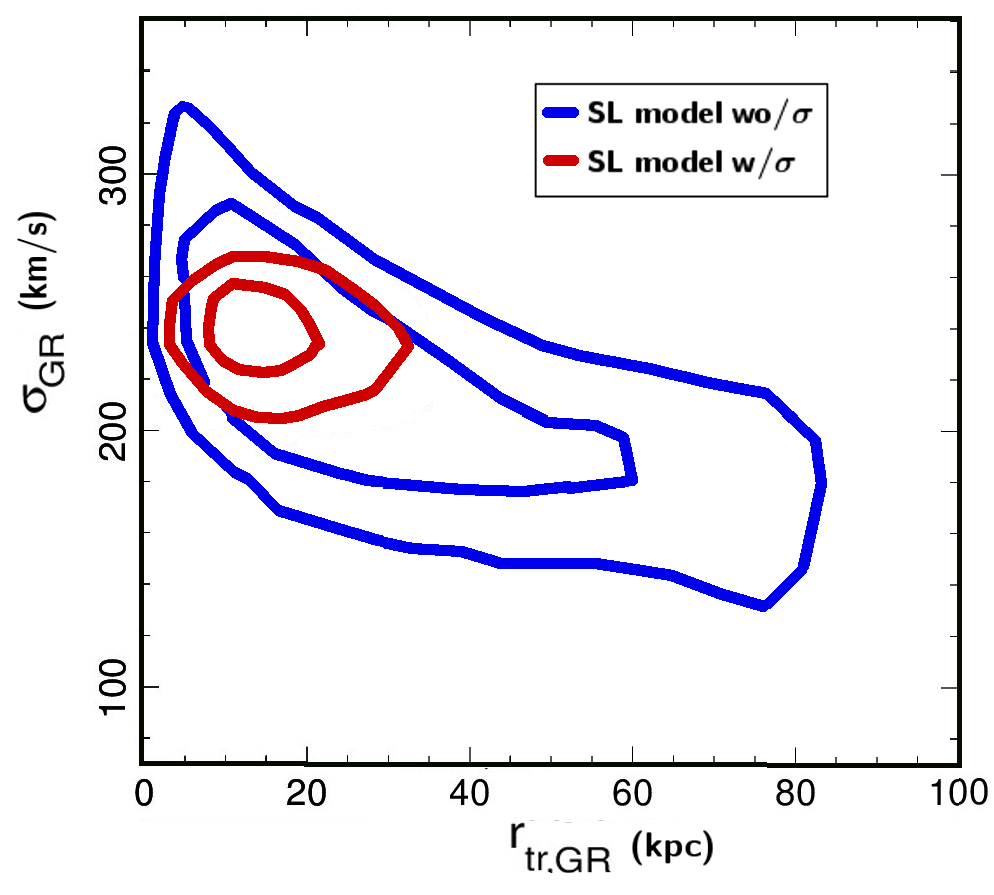}
\caption{\small Probability Contours of the GR velocity dispersions versus the truncation radius from the MCMC sampling for the the model ``$\rm wo/\sigma$'' (in blue) and ``$\rm w/\sigma$'' (in red). The knowledge of the galaxies velocity dispersions improves the constraint on the global scaling relation, tightening the constraints on the galaxy truncation radii by $\sim50\%$.}  
        \label{fig:GR_w_wo}
\end{figure}
The velocity dispersions predicted from SL for the 4 galaxies optimised individually are in good agreement with the measured values, in particular for G1 and G2, which are well constrained through the lensed system 3-4. The velocity dispersion predicted for the reference galaxy, $\rm \sigma_{SL,GR}=214^{+40}_{-32}$\,km/s, is slightly lower than the measured $\sigma_{\rm sp,GR}=233\pm12$\,km/s, but still consistent within the 1$\sigma$ errors.\\ From the comparison of the two pointlike models we find that lensing predictions for galaxies velocity dispersions are overall in good agreement with spectroscopic measurements. We reached a similar results in \citet{Eichner2013}, where the velocity dispersions predicted from strong lensing for cluster members in the core of MACS1206 were in great agreement with the $\sigma$ estimated from the Faber-Jackson relation.   
In Fig.~\ref{fig:GR_w_wo}, we plot the probability contours from the MCMC sampling for the truncation radius and velocity dispersion of the GR. 
The results from the model ``$\rm wo/\sigma$'' show a clear degeneracy between these two parameters (see also Eq.~\ref{eq:m_bbs}), which is broken only in the analysis ``$\rm w/\sigma$''. 
The inclusion of velocity dispersion measurements allows us to improve the constraints on the galaxy sizes by $\sim50\%$ reaching uncertainties of a few kpc on the truncation radii.
The truncation radius scaling relations are 
 \begin{equation}
  r_{\rm tr,``wo/\sigma"}=23.1_{-12.8}^{+29.4} \rm kpc \left(\frac{\sigma}{214_{-32}^{+40}\,km/s}\right)^\frac{4}{3}
  \label{eq:scal_law_nosigma}
 \end{equation}
 \begin{equation}
  r_{\rm tr,``w/\sigma"}=13.2_{-4.3}^{+6.2} \rm kpc \left(\frac{\sigma}{238\pm15\,km/s}\right)^\frac{4}{3}
\label{eq:scal_law_wsigma}
  \end{equation}
for the models ``$\rm wo/\sigma$'' and ``$\rm w/\sigma$'' respectively.  In Fig.~\ref{fig:scaling_law} we plot Eq.\,\ref{eq:scal_law_nosigma} and Eq.\,\ref{eq:scal_law_wsigma} with their 68\% confidence levels. 
In the model ``$\rm wo/\sigma$'', all the galaxies individually optimised lie within the 68\% confidence levels of the scaling relation. \\
When we include the measured velocity dispersions in the analysis, the reference galaxy GR gets a smaller $r_{\rm tr}$ which is better constrained by a factor of 3 to 4. However, we get no improvement on measuring the halo size of the other galaxies individually optimised. 
These galaxies  show a large deviation from the scaling relation.
However, their truncation radii have large errors, such that these galaxies are consistent with the scaling law within $1-2\sigma$ errors.  
The truncation radius of all the other galaxies with measured $\sigma_{\rm sp}$, are scaled with the light according to Eq.~\ref{eq:r_tr}.
They are all in agreement with the scaling relation at the $1\sigma$ level. 
The scaling law ``$\rm w/\sigma$'' is consistent at the 1$\sigma$ level with the law obtained from the  model ``$\rm wo/\sigma$'', and now constraints on the truncation radii are improved by a factor of 50\%.\\ 
The smooth DH parameters  are consistent within the $1\sigma$ errors for both models (see Tab.~\ref{tab:lensing_models}). Including the velocity dispersions helps to constrain more tightly all the DH parameters, except for the PA, where the uncertainty rises by $6\%$.
The external shear is low for both models ($\rm \gamma_{``wo/\sigma"}=0.07\pm0.03$ and $\rm \gamma_{``w/\sigma"}=0.04\pm0.02$)  and  in agreement within the $1\sigma$ errors. 
 In the appendix \ref{appendix} we present the MCMC sampling of the DH parameters for both  models.\\
Strong lensing analyses allow  high precision measurements of the projected mass profile of the lens  within the observed lensing features. 
For both the pointlike models we obtain the same projected mass $M(<50\,\rm kpc)=1.7\pm0.03\times10^{13}M_{\odot}$ enclosed within a radius of 50 kpc, which is the distance of the giant radial arcs (system 1-2) from the cluster centre. This result is in agreement with previous analyses, e.g. with  \citet{Newman2011} and \citet{Zitrin2011}, who find a total projected mass within r=50kpc of  $M(<50\,\rm kpc)=2\times10^{13}M_{\odot}$ and $M(<50\,\rm kpc)=2.2\times10^{13}M_{\odot}$, respectively (both masses are provided without errors). The global models present differences in the mass components parameters due to the different constraints and mass components used in the different analyses, but they show agreement on the total mass predictions probed by strong lensing. 

\section{Surface Brightness reconstruction}
\label{sec:extended_image}
In this section we  perform the surface brightness reconstruction of the southern giant arc corresponding to the lensed systems 3-4 in the pointlike models (see Fig.\,\ref{fig:a383rgb}). 
This is a lensed source at redshift $z_{\mathrm{sp}}=2.58$ which bends between several cluster members. For two of these galaxies, G1 and G2, we have measured velocity dispersions. 
By performing the surface brightness reconstruction of these arcs we aim  to directly measure the truncation radius of these two cluster members, which are the only unknown parameters of the profiles adopted to describe their mass.\\
To perform the surface brightness reconstruction, \texttt{GLEE} uses a linear inversion method \citep[see][]{Warren2003}.
It reconstructs the pixellated brightness distribution of the source, with regularisation of the source intensity through a  Bayesian analysis \citep[see][for a detailed description of this technique]{Suyu2006}.\\
We reconstruct  system 3-4 in the HST/ACS/F775W filter, in which the arcs are bright and at the same time the light contamination from the close cluster members is still low. 
In order to reconstruct only the light from the arcs and avoid contamination from nearby galaxies, we subtract the galaxies close to  system 3-4 using the SNUC\footnote{see http://astronomy.nmsu.edu/holtz/xvista/index.html and \citealt{Lauer1986}} isophote fitting routine, which is part of the XVISTA image processing system. 
Within CLASH, we apply SNUC to derive two-dimensional models of  early-type galaxies in the CLASH clusters since it is capable of simultaneously obtaining the best non-linear
least-squares fits to the two-dimensional surface brightness distributions in multiple, overlapping galaxies (see \citealt{Lauer1986}). \\
\begin{figure*}
\centering
\includegraphics[width=18cm]{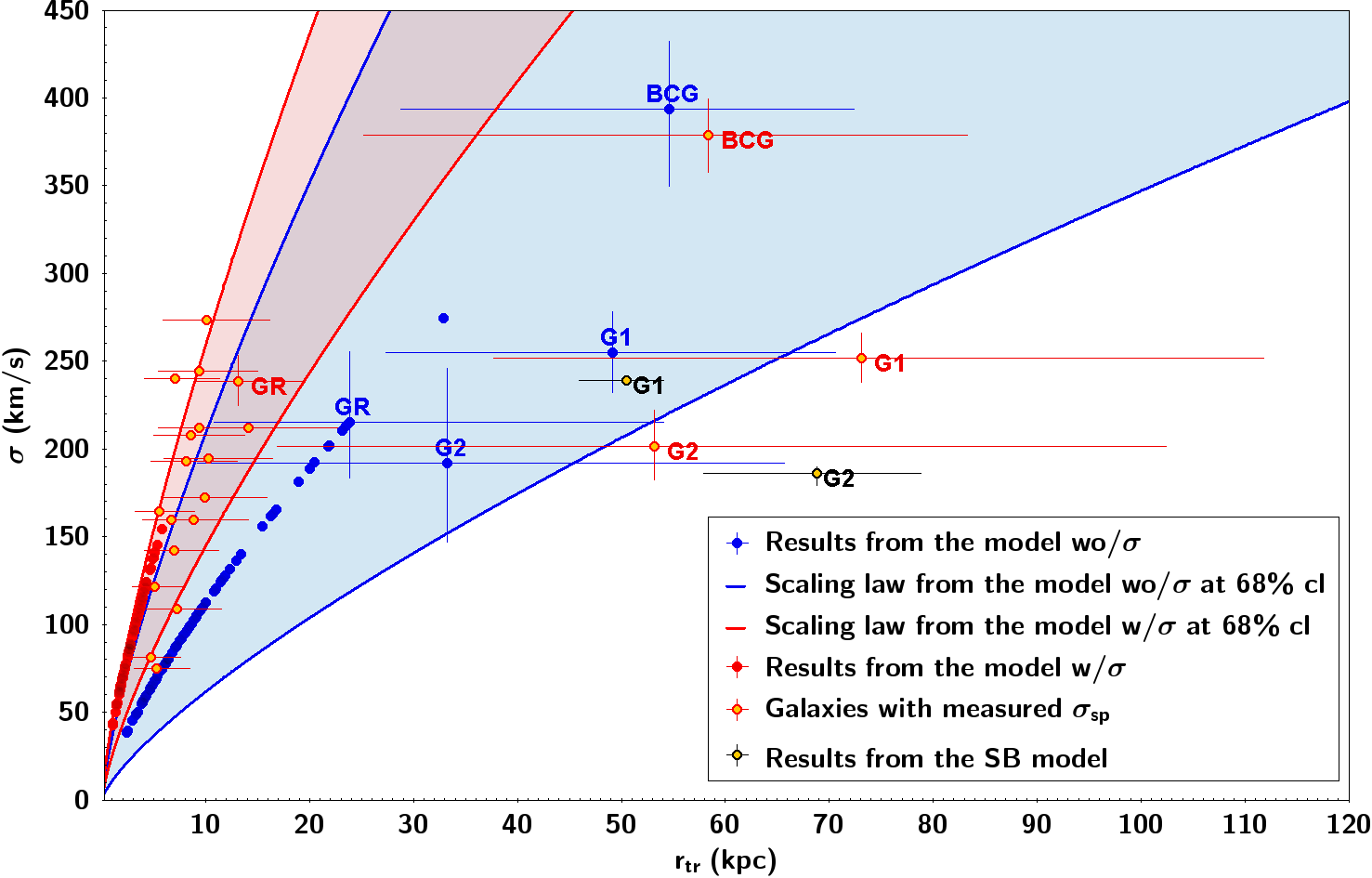}
\caption{\small Galaxy scaling relations for the models ``$\rm wo/\sigma$'' and ``$\rm w/\sigma$''. The blue lines are the 68\% confidence level for the scaling law from the model ``$\rm wo/\sigma$''. All the cluster members (blue circles) are scaled according to this relation, except the 4 galaxies which we optimise individually and which are labelled with their respective ID. 
The 68\% confidence levels of the scaling relation obtained from the model ``$\rm w/\sigma$'' is shown in red, and the red circles are the cluster members scaled according to this relation. We plot in yellow the galaxies with measured velocity dispersion, for which the velocity dispersions are fixed to the measured $\sigma_{\rm sp}$ and only the truncation radii are scaled according to the galaxy luminosity. These galaxies scatter around the cluster $\sigma$-$r_{\rm tr}$ scaling relation from the model ``$\rm w/\sigma$'', but they are all consistent with this scaling law  within the 1$\sigma$ errors. The galaxies individually optimised, BCG, G1 and G2, have a large scatter from the luminosity relation, and 
in particular G1 and G2 present truncation radii larger than the 68\% cl of the scaling law ``$\rm w/\sigma$''. In addition we plot in black the prediction for G1 and G2  obtained from the surface brightness reconstruction of the southern giant arc (see Sec.\ref{sec:extended_image}). Also in this case these galaxies present a large deviation from the scaling relations, but the results from the three analysis are all consistent within each other for both galaxies.
}  
\label{fig:scaling_law}
\end{figure*}

\begin{figure}
\centering
\includegraphics[width=8.5cm]{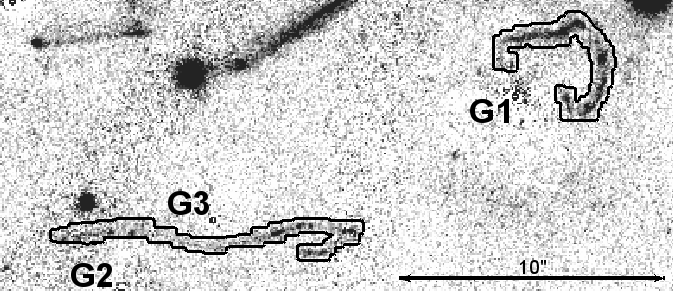}
\caption{\small Cutout of the system 3-4 in the HST/ACS/F775W filter. In this images the three galaxies close to the system,  G1, G2 and G3, are subtracted using the SNUC routine. In black we trace the contours of the area we reconstruct in the surface brightness reconstruction of this system. }  
        \label{fig:arc}
\end{figure}

We perform the surface brightness reconstruction of the arcs where $S/N > 0.5$. In Fig.\,\ref{fig:arc} we show the arcs in the F775W filter with the bright nearby galaxies subtracted, and we show  in black the contours of the area we mask for reconstruction. 
When performing the surface brightness reconstruction of systems 3-4, we fix the mass profile parameters of the smooth dark halo, GR and BCG to the values obtained from the model ``$\rm w/\sigma$''. 
Then we only optimise the mass profile parameters associated with the three cluster members G1, G2 and G3 close to the arcs (see Fig.\,\ref{fig:arc} ).  
As before, position and shape parameters of these three galaxies are estimated using the values traced by the light. 
For G1 and G2 we optimise the PA and b/a with gaussian prior using their 10\% error as  width. 
We also optimise their  measured velocity dispersions  within their uncertainties using a gaussian prior. 
For G3  we have no measured $\sigma_{\rm sp}$, so we use the $\sigma_{\rm G3}$ resulting from the model ``$\rm w/\sigma$'', and optimise it within the 1$\sigma$ uncertainties  with a gaussian prior.  
Finally we optimise the truncation radii of these three galaxies (G1 to G3) within the wide range of [1,100]\,kpc with a flat prior and we also allow for a core radius for G1 and G2. \\ 

\begin{figure}
\centering
\includegraphics[width=8.5cm]{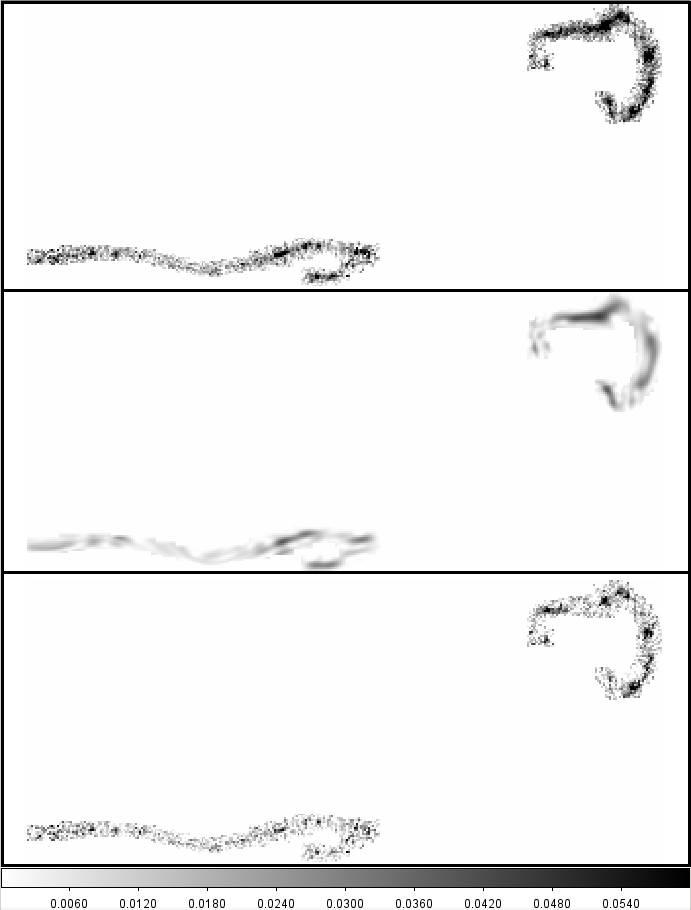}
\caption{\small Surface brightness reconstruction of the giant southern arcs ($20\times10$ arcsec cutout, which corresponds to $\sim60\times30$ kpc at the cluster redshift). Upper panel show the arcs in the HST/ACS/F775W filter, the central panel shows the  reconstruction of the arc in this filter, and the lower panel shows the residuals. }  
        \label{fig:esource_im}
\end{figure}

\begin{figure}
\centering
\includegraphics[width=8cm]{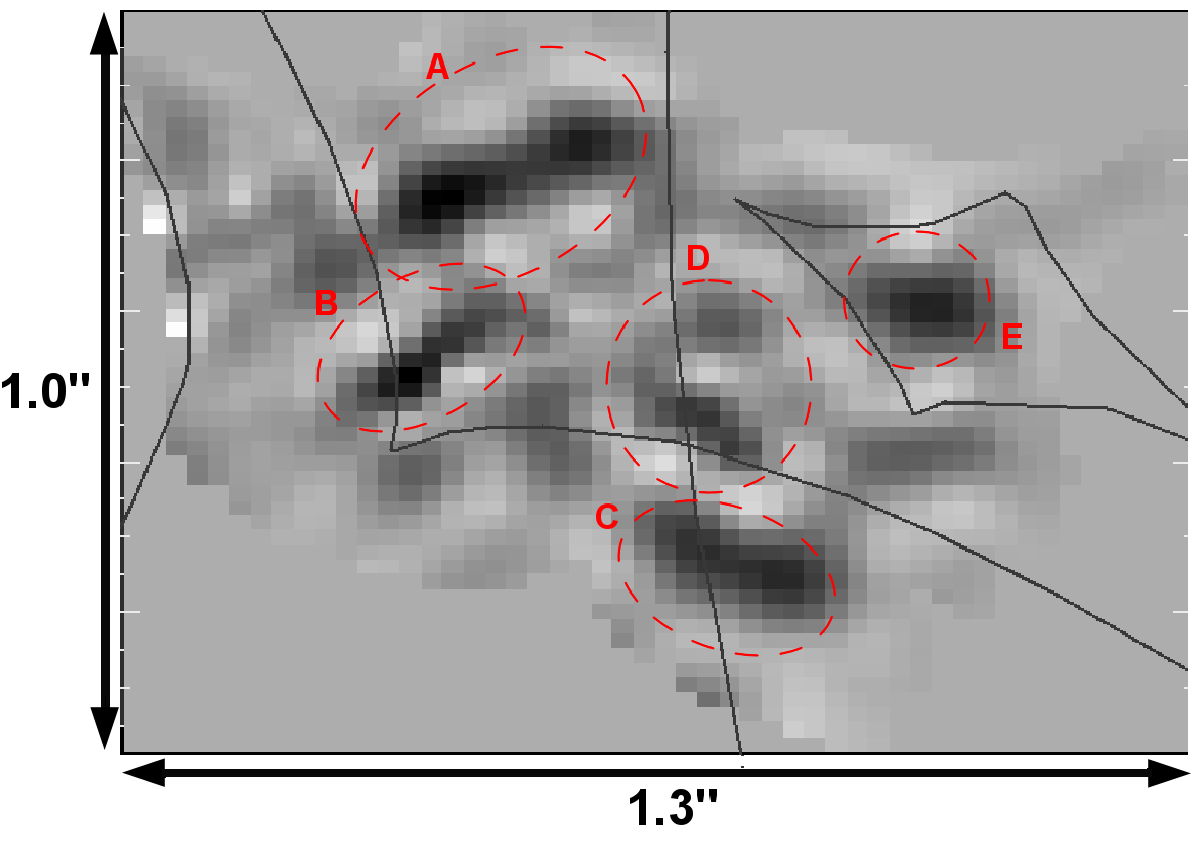}
\caption{\small Source reconstruction of the southern arc ($1.3\times1.0$ arcsec cutout, which corresponds to $\sim10\times8$ kpc at the redshift of the source). The gray lines are the caustics. The reconstructed source is composed of 5 clumps (red dashed contours). The clumps A-B corresponds to the system 3 in the pointlike models, while C-D-E are the light reconstruction of system 4.}  
        \label{fig:esource_sr}
\end{figure}

The final best model has a reduced $\chi^2_{\rm img}=1.4$ from all images positions.
In Fig.\,\ref{fig:esource_im} we show the arc reconstructed, the original image, and the residual between these two images. 
The $\chi^2$ from the pixellated surface brightness reconstruction of the southern arcs is $\chi^2_{\rm SB}=0.78$.
In Fig\,\ref{fig:esource_sr} we present the reconstruction of the unlensed source. 
It shows an irregular light distribution which consists of 5 clumps. 
The clumps A-B corresponds to the system 3 in the pointlike models, while the clumps C-D-E to system 4. 
Irregular light distribution seems to be common to galaxies at redshift $z>2$. 
The Hubble morphological sequence applies to galaxy population from the local Universe up to intermediate redshifts $z\sim1-2$ \cite[e.g., see][]{Glazebrook1995,Stanford2004}.
At higher redshifts the majority of galaxies shows irregular and clumpy morphology \cite[e.g., see][]{Dickinson2000,Conselice2005,Talia2014}. 
The source reconstructed has a size of $\sim0.5\arcsec$, which corresponds to $\sim4$\,kpc at $z_s=2.58$. 
Galaxies in the redshift range  $z\sim2-3.5$  typically have radius ranging between 1-5 kpc \cite[e.g., see][]{Bouwens2004,Oesch2009}. Thus the size of the source we reconstruct at $z_s=2.58$ is consistent with value expected for galaxies at high redshift. \\


In Table \ref{tab:masses} we list the most probable mass parameters and their respective $1\sigma$ uncertainties for  G1, G2 and G3. 
Here we focus on the parameters for G1 and G2, to compare them with results from the pointlike models.  
The ellipticity and PA are stable relative to the values extracted from the light profiles for both galaxies. 
G1 gets a velocity dispersion of $\sigma=239\pm2$\,km/s, which is consistent with previous results within the $1\sigma$ uncertainties. 
The core radius is $1.3\pm0.1$\,kpc and the truncation radius is $50.5^{+3.6}_{-4.6}$\,kpc. 
For G2 we get $\sigma=186^{+4}_{-7}$\,km/s, $r_{c}=0.3_{-0.2}^{+0.3}$\,kpc (consistent with zero) and $r_{\rm tr}=68.8_{-10.9}^{+10.0}$\,kpc. 
In Fig.~\ref{fig:scaling_law} we plot the results for G1 and G2 to compare them with the prediction from the scaling relations obtained from the pointlike modelling. 
In the previous models these two galaxies get truncation radii which are  several times larger than the predictions from the respective luminosity scaling laws in Eq.~\ref{eq:scal_law_nosigma} and \ref{eq:scal_law_wsigma}. 
The surface brightness reconstruction of the southern arcs leads to similar results. Both the  galaxies have  truncation radius larger than the respective predictions from the global scaling law. 
However, comparing the $r_{\rm tr}$ prediction of these two galaxies from the three analyses performed in this work, they  are all consistent with each others within the $1\sigma$ errors. 
In the Appendix~\ref{appendix} we plot the Monte Carlo Markov Chain sampling of the dark halo parameters for both G1 and G2. 
The total masses associated with the two galaxies (see Eq.~\ref{eq:m_bbs1}) are $M_{G1}=2.1\pm0.2\times 10^{12}M_\odot$ and $M_{G2}=1.7\pm1.3\times10^{12}M_\odot$, which are consistent with the mass estimations from the pointlike models.  
See Tab.~\ref{tab:masses} for a summary of the mass profile parameters.\\
To infer the amount of stripped dark matter for galaxies in cluster cores we can estimate the total mass that  G1 and G2 would have if they were in underdense environments, and compare them to  their total mass estimated with lensing in the cluster core. 
\cite{Brimioulle2013} estimated the $r_{\rm tr}$-$\sigma$ scaling law for early type field galaxies, getting  $\mathrm{r_{tr,field}}=~245^{+64}_{-52}h^{-1}_{100}$kpc for a reference galaxy with $\sigma=~144$km/s, assuming that $r_{\rm tr}\propto\sigma^2$ in fields. Using this relation and  Eq.\,\ref{eq:m_bbs1} we can derive the mass that G1 and G2 would have in the field.   
Assuming that the velocity dispersion of the halo does not change when a galaxy infalls in cluster and during the stripping process, we get that $\rm M_{tot,SL}^{G1}/M_{tot,fields}^{G1}=0.07$ and   $\rm M_{tot,SL}^{G2}/M_{tot,fields}^{G2}=0.17$, which imply that 93\% and 83\% of the mass has been stripped respectively for G1 and G2. This results is in agreement with numerical simulations of tidal stripping processes \cite[see][]{Warnick2008,Limousin2009} which estimate that $\sim90\%$ of the mass is lost for galaxies in cluster cores.

\begin{table} 
\caption{Most probable parameters with the respective $1\sigma$ errors for the dPIE mass distribution of the cluster members close to the reconstructed giant arcs. The total mass is estimated according to Eq.\,\ref{eq:m_bbs}}
\centering
\footnotesize
\begin{tabular}{|c|c|c|c|}
\hline
\hline
Galaxy& $\rm ``wo/\sigma"$& $\rm ``w/\sigma"$& $\rm Ext\_model$\\
\hline
\hline
%BCG & $ 0.65$& $ 60.5$ &$ 53.0$  &  $64.3$    \\
%\hline
%RG & $ 0.48$& $ 7.7$ &$ 5.1$  &  $2.7$    \\
%\hline
\textbf{G1} & & & \\
\hline
b/a & 0.8& 0.8 &  $0.82\pm0.01$ \\
PA & 151 & 151 & $150.7\pm0.6$\\
$\sigma$ [km/s] & $253\pm23$&$252\pm14$ & $239\pm2$\\
$r_{core}$ [kpc]& 0. & 0.& $1.3\pm0.1$\\
$r_{\rm tr}$ [kpc]& $47.8\pm20.9$&$56.8^{+24.8}_{-25.6}$ & $50.5_{-4.6}^{+3.6}$\\
$M_{}\,[10^{12}M_\odot]$& $ 2.2\pm1.4$ &$ 2.6_{-1.5}^{+1.4}$  &  $2.1\pm0.2$  \\ 
\hline
\textbf{G2}& & & \\
\hline
b/a & 0.58& 0.58 &  $0.57\pm0.01$ \\
PA & 63 & 63 & $63\pm1$\\
$\sigma$ [km/s] & $194^{+54}_{-45}$& $201\pm20$ & $186_{-7}^{+4}$\\
$r_{core}$ [kpc]& 0. & 0.& $0.3_{-0.2}^{+0.3}$\\
$r_{\rm tr}$ [kpc]& $32.2^{+31.5}_{-23.4}$ &$53.2^{+49.2}_{-36.3}$ & $68.8^{+10.0}_{-10.9}$\\
$M_{\rm tot}\,[10^{12}M_\odot]$& $ 0.9^{+1.3}_{-1.1}$ &$ 1.6_{-1.4}^{+1.8}$  &  $1.7\pm1.3$  \\
\hline
\textbf{G3}& & & \\
\hline
b/a & 0.93& 0.93 &  $0.57\pm0.01$ \\
PA & 65 & 65 & $65$\\
$\sigma$ [km/s] & $109^{+20}_{-16}$& $120\pm7$ & $128\pm2$\\
$r_{core}$ [kpc]& 0. & 0.& $0.$\\
$r_{\rm tr}$ [kpc]& $9.6^{+12.2}_{-5.3}$ &$4.1^{+2.5}_{-1.7}$ & $2.9\pm0.4$\\
$M_{\rm tot}\,[10^{12}M_\odot]$& $ 0.08^{+0.14}_{-0.07}$ &$ 0.04_{-0.03}^{+0.02}$  &  $0.04\pm0.01$  \\
\hline
\hline
\end{tabular}
\label{tab:masses}
\end{table}

\section{Discussion and Conclusions}
\label{sec:conclusions}
In this paper we measured the mass distribution in the core of A383  using pointlike lensing constraints and by reconstructing the surface brightness distribution of giant arcs.
For the first time we include in the lensing analyses the measurements of velocity dispersions for 21 cluster members.
These allow us to refine individually  the constraints on the galaxy mass component and  on the smooth dark halo mass profile. \\
In Sec.\,\ref{sec:pointlike_model} we  constructed two parallel models, one in which we include the measured $\sigma_{\rm sp}$ and the other in which do not use such information.
We find that the $\sigma_{\rm SL}$ values are globally in agreement with the measured values at the $1\sigma$ level (see Fig.\,\ref{fig:sigma_sl_sp}). Only few galaxies have  $\sigma_{\rm SL}$ slightly different from the measured velocity dispersions, which are faint galaxies with large errors on the measured velocity dispersions. However they agree at the $2\sigma$ level with the spectroscopic measurements.\\ 
In particular,  when we optimise the mass profiles of cluster members individually, taking advantage of stronger constraints from lensing, the $\sigma_{\rm SL}$ predictions are in great agreement with the $\sigma_{\rm sp}$ measurements. 

The galaxy chosen as reference for the luminosity scaling relations has measured velocity dispersion. Thus, once we fix the exponents of the scaling relations (Eq.~\ref{eq:F_J} and Eq.~\ref{eq:r_tr}), the only parameter we need to constrain to estimate the global scaling laws is the truncation radius $r_{\rm tr,GR}$. 
The results of the pointlike models show that the knowledge of the cluster members velocity dispersions allows to improve the constraints on  the $r_{\rm tr,GR}$ and on the scaling relation by 50\%. 
\citet{Faber2007} investigated the luminosity function for red and blue galaxies in several redshift bins up to $z\sim1$.
For the red galaxy sample with redshift $0.2\leqq z<0.4$, a typical $L^*$ galaxy has  $M^{*}_{B}=-20.95^{+0.16}_{-0.17}$ in AB system. 
According to our final scaling relation $``w/\sigma"$ (Eq.~\ref{eq:scal_law_wsigma}), such a typical $L^*$ red galaxy at $z\sim0.2$ has a truncation radius of $r_{\rm tr}^{*}=20.5^{+9.6}_{-6.7}$\,kpc,  velocity dispersion of $\sigma_{*}=324\pm17$\, km/s and total mass $M_{\rm tot}^*=1.57_{-0.54}^{+0.75}\times10^{12}\rm M_{\odot}$.\\
\cite{Nat2009}, combining strong and weak lensing analyses, investigated the dark halo of galaxies in the core of CL 0024+16 at $z=0.39$ for early and late type galaxies as a function of their distance from the cluster centre. 
They found that the dark halo mass of a fiducial $L^*$ early type galaxy increases with the distance from the cluster centre, from  $M^*=6.3_{-2.0}^{+2.7}\times 10^{11}\,M_{\odot}$ in the core ($r<0.6$\,Mpc) to $M^*=3.7_{-1.1}^{+1.4}\times 10^{12}\,M_{\odot}$ in the outskirts. 
This is  consistent with our results for a $L^*$ galaxy in the core of A383 at the $2\sigma$ level.\\
\cite{Limousin2009}, using N-boby hydrodynamical simulations, probed the tidal stripping of galaxy dark halos in clusters in the redshift range $z_{\rm cl}=[0,0.7]$.
They used the half mass radii $r_{1/2}$ of galaxies to quantify the extent of their dark halos, which correspond to our $r_{\rm tr}$ for dPIE profile with vanishing core radius. 
They found that the $r_{1/2}$ and the total dark halo mass of the galaxies decrease moving from the outskirts to the core of the clusters, showing that galaxies in the core experience stronger stripping than the ones in the outer regions. In particular, galaxies in the core ($r<0.6$\,Mpc) are expected to have $r_{1/2}<20$\,kpc. In this work we analysed the halo properties of galaxies in the core of A383, with projected radial distance $R<1.5'=0.3$\,Mpc. Our results from the model $''w/\sigma"$ are in great agreement with the predictions of \citet{Limousin2009}, confirming that the sample of cluster members we investigated in the core of A383 experienced strong tidal stripping. \\

\begin{figure*}
\centering
\includegraphics[width=18cm]{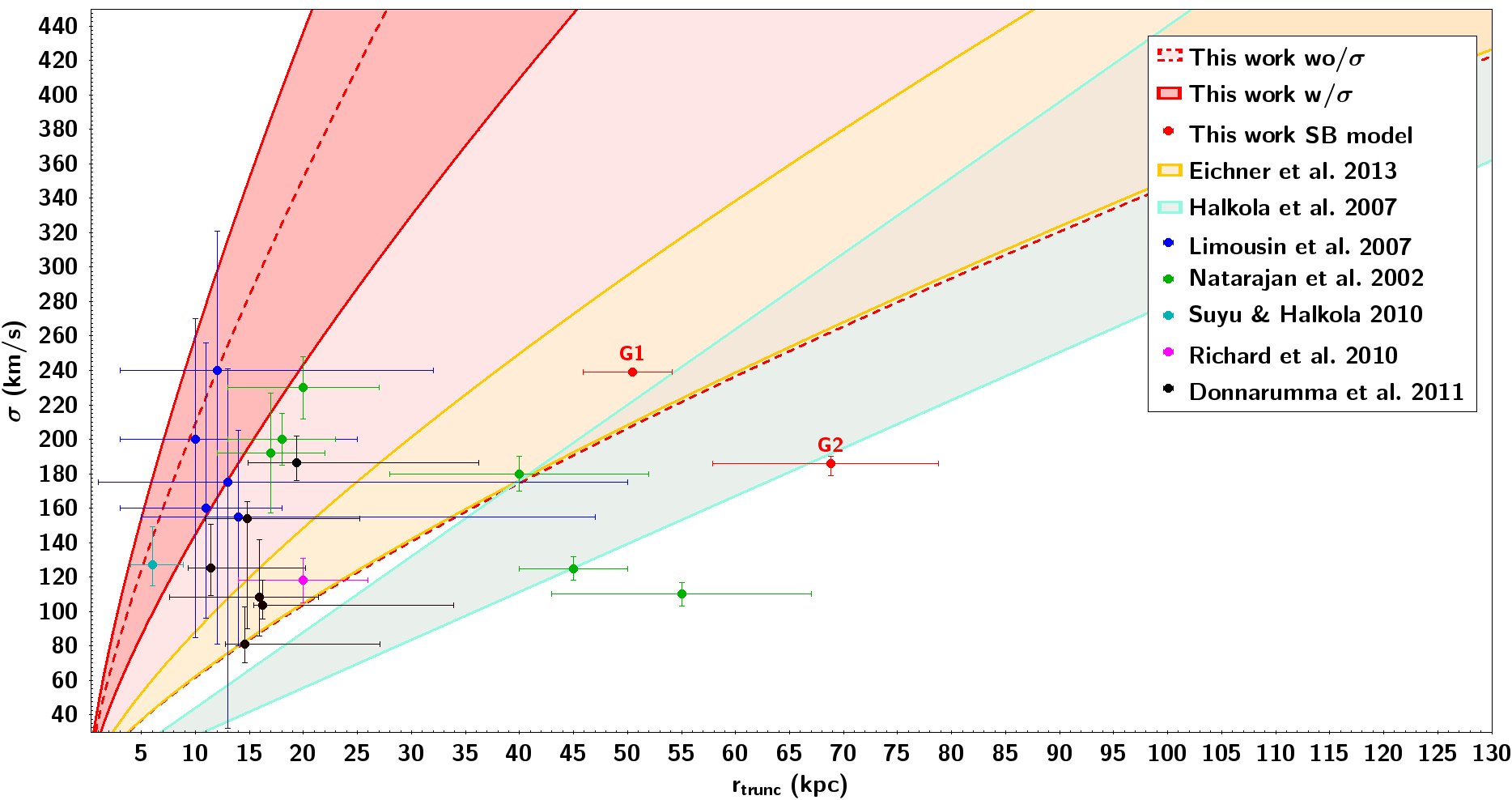}
\caption{\small Galaxy scaling relations  resulting from our and previous lensing analyses.
The  red lines are the 68\% confidence levels of the  scaling relations obtained in this work from the model ``$\rm wo/\sigma$'' (red dashed lines) and ``$\rm w/\sigma$'' (red lines). We also plot the measurements of the dark halo sizes of G1 and G2 resulting from the surface brightness reconstruction of the southern giant arcs. In yellow and cyan we plot the 68\% confidence levels of the scaling relations derived in \citet{Eichner2013} and \citet{Halkola2007} respectively. \citet{Nat2002} and \citet{Limousin2007} derive the halo sizes of galaxies in the core of several clusters, using  respectively strong and weak lensing analyses. Their results are plotted as green (\citet{Nat2002}) and blue circles (\citet{Limousin2007}). The uncertainties on the measurements of \citet{Limousin2007} are 1- 2- and 3- $\sigma$ errors, depending on the cluster \citep[see][ for details]{Limousin2007}.  In addition we plot individual measurements of galaxy halos in a group, from \citet{Suyu2010} (plotted in cyan), and in clusters, from  \citet{Richard2010} (in magenta) and \citet{Donnarumma2011} (in black). The measurements from \citet{Donnarumma2011} are the parameters for the six galaxies individually optimised in the core of A 611, corresponding to the ``case 6'' presented in their paper.
Our scaling relation from the modelling ``$\rm wo/\sigma$'' is in agreement with all the previous measurements, except for two clusters from \citet{Nat2002}, while the tighter relation resulting from our model ``$\rm w/\sigma$'' is consistent with the results from \citet{Limousin2007}, \citet{Suyu2010} and part of results from \citet{Nat2002}. See text for more details. 
}  
        \label{fig:scaling_law_lit}
\end{figure*}

In Fig.\,\ref{fig:scaling_law_lit} we compare our results for the scaling law between the truncation radius and velocity dispersion with results from previous analyses. 
In \cite{Eichner2013} we measured the galaxies' scaling relation in the cluster MACS1206 at $z=0.439$ performing an analysis similar to the one presented here for A383, but without the knowledge of cluster members' velocity dispersions. 
For MACS1206  we  obtained $r_{\rm tr}=35\pm8\,\rm kpc (\sigma/186\,km/s)^{4/3}$, which is consistent with the result for A383 from the pointlike model ``$\rm wo/\sigma$'', but it is not in agreement at the 1$\sigma$ level with the tighter scaling relation we get from the model ``$\rm w/\sigma$''. 
This is also the case when we compare our results with the ones presented in \cite{Halkola2007}, where strong lensing is used to derive the size of galaxy halos in the core of Abell 1689.   
They tested the assumption of two different exponents for the  $r_{\rm tr}$-$\sigma$ relation, using $\alpha/\delta=1,2$ (see Eq.\,\ref{eq:r_tr}). 
The reference truncation radii resulting from their two models are consistent  and they conclude that galaxies in the core of the cluster are strongly truncated.
For simplicity in Fig.\,\ref{fig:scaling_law_lit} we plot only their results for   $\alpha/\delta=1$, which is closer to the exponent assumed in our analyses.     
The scaling relations from \citet{Eichner2013} and \cite{Halkola2007} deviate from our relations.
This can be a result of the different clusters analysed.
Another reason could be that, by scaling all the cluster members (including the brighter ones) with the same law, the resulting sizes are overestimated. 
Bright cluster members, which have been central galaxies before accretion to the cluster, have not yet been strongly stripped as fainter galaxies which have been satellites for a long time.
Indeed one expects that the dispersion of halo mass is larger for bright galaxies than for fainter ones, depending on whether they have been a satellite or central galaxy at accretion of the cluster. 
In our analysis several brighter central galaxies (GR, BCG, G1, G2) are individually optimised, and the scaling laws mainly applies to galaxies which have been satellites for a long time.\\ 
\cite{Suyu2010}  derived the size for a satellite halo in a galaxy group at $z=0.35$, which has a projected distance from the centre of galaxy group of $R\sim26$\,kpc. The truncation radii and velocity dispersion estimated for this satellite are $r_{\rm tr}=6.0_{-2.0}^{+2.9}$\,kpc for $\sigma_{sat}=127_{-12}^{+22}$\,km/s respectively. This  is in good agreement with predictions from our scaling law $``w/\sigma"$ at low velocity dispersions, and support that our scaling law is representative for satellite galaxies. \\
\cite{Nat2002}, combining strong  and weak lensing analyses, investigated properties of galaxies  in  6 massive clusters spanning the redshift range $z=0.17-0.58$, using archival HST data. 
They found that galaxies are  tidally truncated in clusters, and in particular their results for 3 clusters of the sample (A2390, AC114, CL0054-27) are in good agreement with our results from the modelling ``$\rm w/\sigma$''.\\ 
\cite{Limousin2007} used weak lensing to measure the size of galaxies in 5 clusters at $z\sim0.2$, including A383, covering a wide FOV with $R<2$\,Mpc. 
Globally they find that  galaxies with velocity dispersion within $[150,250]$\,km/s have truncation radii lower than 50 kpc, with mean value of 13 kpc, which is consistent with predictions from our scaling laws. In particular for A383 they predicted $\rm r_{\rm tr}=13_{-12}^{+37}$\,kpc for a galaxy with $\sigma=175_{-143}^{+66}$\,km/s  (in agreement with our results).  \\ 
Finally \cite{Richard2010} and \cite{Donnarumma2011} measured the halo size of individual galaxies in the core of Abell 370 ($z=0.375$) and Abell 611 ($z=0.288$) respectively, taking advantage of direct strong lensing constraints on the galaxies. 
Their analyses predict larger truncation radii for these galaxies when compared to our ``$\rm w/\sigma$'' scaling law, but they are consistent with our results from the model ``$\rm wo/\sigma$'' at the $1\sigma$ level (see Fig.\,\ref{fig:scaling_law_lit}). \\
The estimates  from these previous works are still degenerate with the velocity dispersions they used. 
Here in this work, for the first time we broke this degeneracy using measurements of cluster members velocity dispersions.
\\
To improve the constraints on the halo size of individual galaxies in the core of the cluster, we performed the  surface brightness reconstruction of the southern giant arcs. 
This allowed us to measure the $r_{\rm tr}$ of two close cluster members, G1 and G2, for which we have measured velocity dispersions.
With this analysis we improve the constraints by more than 30\% on the halo  size of these two galaxies. 
The results are also plotted in Fig.\,\ref{fig:scaling_law_lit}, which shows that these two galaxies deviate from the global scaling law derived for the cluster. 
This could mean that G1 and G2 have been central galaxies before accretion to the cluster and  suffered less stripping than fainter galaxies which have been satellites.
However, using Eq.\,\ref{eq:m_bbs} we estimated the total mass associated with the dark halo for G1 and G2 and compared these values with the mass they would have without suffering any stripping for the interaction with the cluster dark halo and the other galaxies. 
It results that 93\% and 83\% of the mass has been  stripped respectively for G1 and G2, in agreement with results from numerical simulations which predict that galaxies in cluster cores loose 90\% of their mass due to tidal stripping. \\

In this paper we have shown  that the degeneracy in the analytic scaling relation, adopted for cluster members in lens modelling, can be broken using measured velocity dispersions of individual cluster galaxies. 
The knowledge of cluster members $\sigma_{\rm sp}$ yields to improvements both on the fit and on the constraints on the mass shape and composition. 
We found that galaxies in cluster cores are strongly truncated, which is overall in agreement with previous measurements and also with prediction from numerical simulations. 
High resolution photometric and spectroscopic data, combined with galaxy kinematics, allow us to constrain to a higher level the galaxy scaling law in core of clusters, and also to individually identify cluster members which deviate from the global scaling law measured for the cluster, as G1 and G2.\\   
This was a first case study on a well studied lensing cluster, A383. 
In the near future we plan to apply this new technique to a larger sample of clusters, and explore further the treasury of using cluster members measured velocity dispersions in lensing analysis.

\section*{Acknowledgements}
This work is supported by the Transregional Collaborative Research Centre TRR 33 - The
Dark Universe and the DFG cluster of excellence ``Origin and Structure of the Universe". 
The CLASH Multi-Cycle Treasury Program (GO-12065) is based on observations made with the NASA/ESA Hubble Space Telescope. The Space Telescope Science Institute is operated by the Association of Universities for Research in Astronomy, Inc. under NASA contract NAS 5-26555. The Dark Cosmology Centre is funded by the DNRF. Support for AZ is provided by NASA through Hubble Fellowship grant \#HST-HF-51334.01-A awarded by STScI. The Smithsonian Institution supports the research of DGF, MJG, and HSH.
We thank Daniel Gruen for his contribution to the improvement of the text.  
\addcontentsline{toc}{chapter}{Bibliography}
\bibliographystyle{mn2efix}
\bibliography{monna_a383}

\appendix
\section{MCMC Sampling of the pointlike models}
\label{appendix}

\begin{figure*}
\centering
\includegraphics[width=10.5cm]{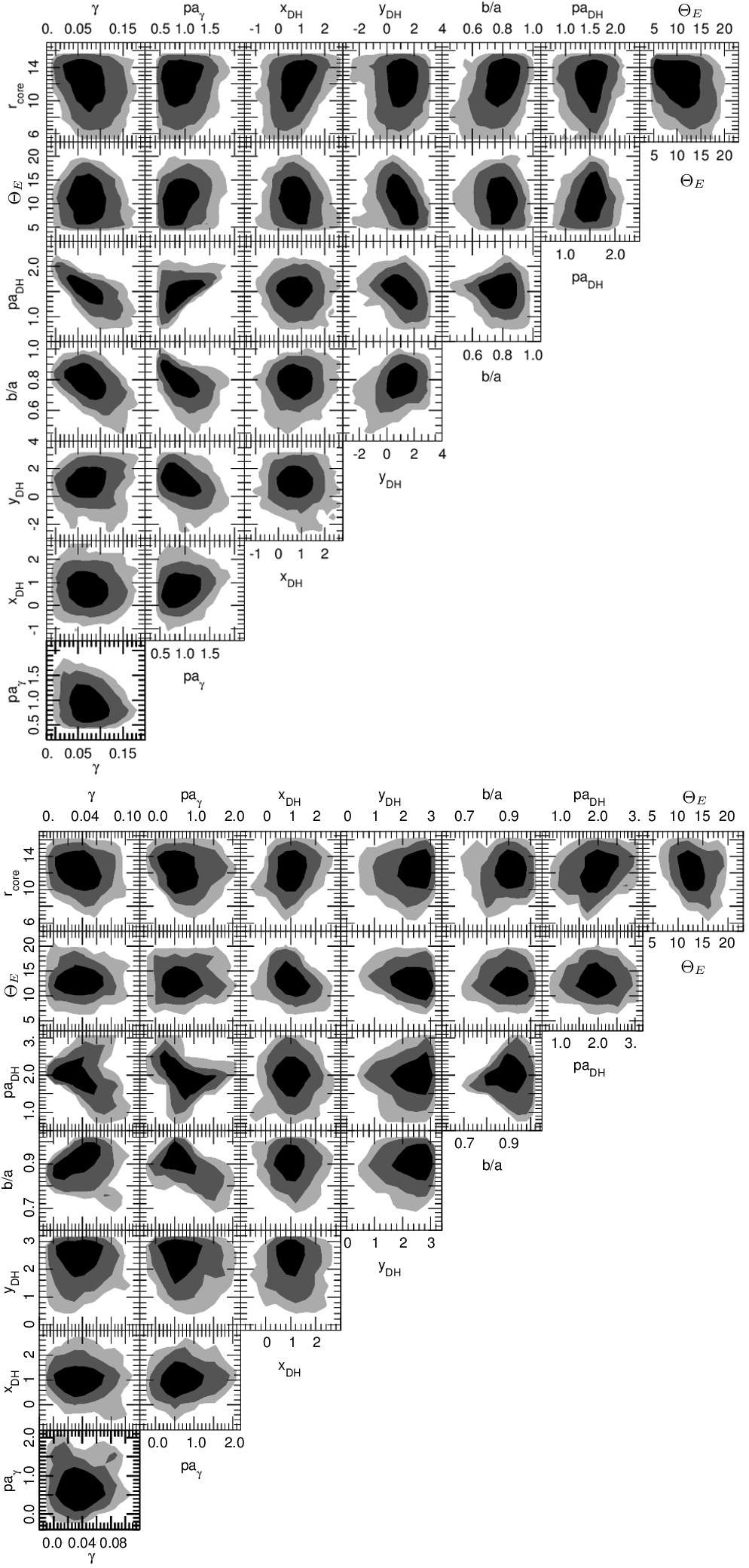}
\caption{\small MCMC sampling of the  DH parameters for the models ``$\rm wo/\sigma$'' (upper panel) and ``$\rm w/\sigma$'' (lower panel). The angles pa$_\gamma$ and pa$_{\rm DH}$ are given in radiants, x$_{\rm DH}$ and y$_{\rm DH}$ in arcseconds with respect to the BCG position, $r_{\rm core}$  and $\Theta_E$ are in arcseconds as well.} 
        \label{fig:mcmc_dh_sig}
\end{figure*}

\begin{figure*}
\centering
\includegraphics[width=10.5cm]{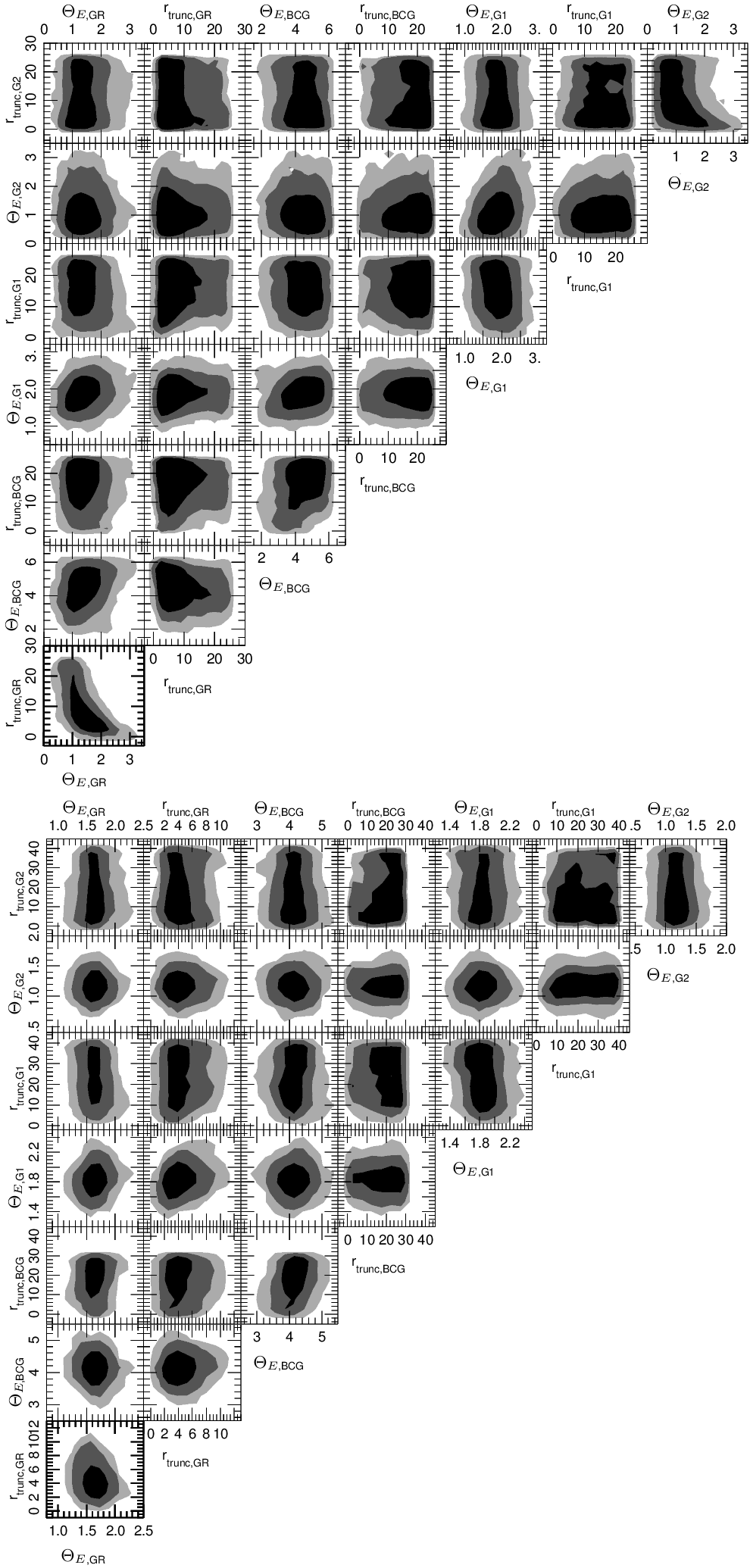}
\caption{\small MCMC  parameters for GR, BCG, G1 and G2 from the model without (upper panel) and with sigma (lower panel). The truncation radii and $\Theta_E$ are in arcseconds as in the previous plot. }  
        \label{fig:mcmc_gal_no_sig}
\end{figure*}

\begin{figure*}
\centering
\includegraphics[width=14cm]{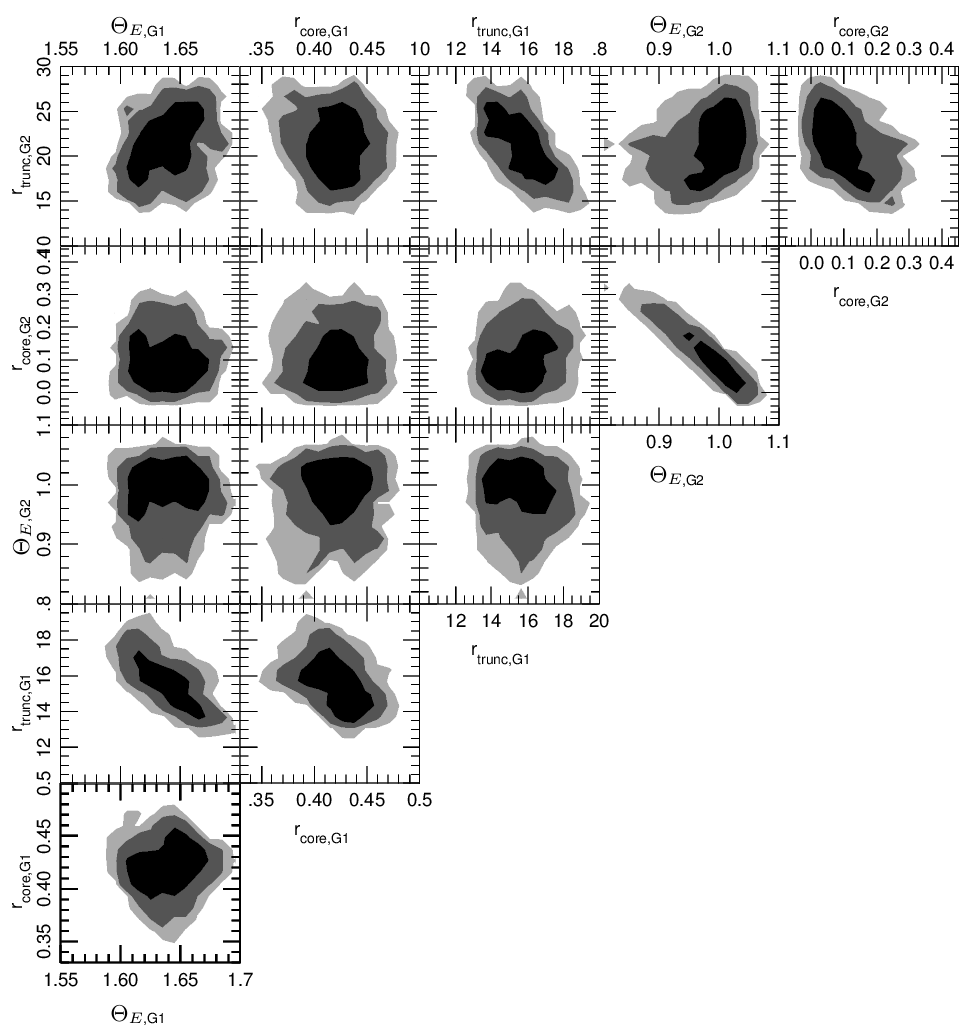}
\caption{\small MCMC sampling of the parameters describing the DH mass profiles of G1 and G2 resulting from the surface brightness reconstruction. The truncation and core radii and $\Theta_E$ are give in arcseconds as in the previous plots.}  
        \label{fig:sb_mcmc}
\end{figure*}

We show here the Monte Carlo Markov Chain sampling of the parameters describing the cluster dark halo and the physical parameters for the galaxies that we optimised individually through our analyses. The gray scales correspond to 68.3\%  (black), 95.5\% (dark gray) and 99.7\% (light gray).
In Fig.\,\ref{fig:mcmc_dh_sig} we show the MCMC sampling of the parameters of the smooth dark halo mass profile of the cluster, both for the pointlike model ``$\rm wo/\sigma$'' (upper panel) and ``$\rm w/\sigma$''. 
In Fig.\,\ref{fig:mcmc_gal_no_sig}  we show the sampling of the mass parameters for the 4 galaxies individually optimised in the pointlike models. The upper panel shows the results for the model ``$\rm wo/\sigma$'' and the lower one the results for the model ``$\rm w/\sigma$''. The galaxies truncation radii present large errors in the model ``$\rm wo/\sigma$'', and including the measured velocity dispersions of the 21 cluster members allows us to improve the constraints on the halo size of the reference galaxy by 50\%.
Performing the surface brightness reconstruction of the southern arcs improves the constraints also on the individual galaxies G1 and G2 close to the arcs, as can be seen from the MCMC sampling of this model presented in Fig.\,\ref{fig:sb_mcmc}.   


\bsp

\label{lastpage}

\end{document}